\documentclass[a4paper,12pt]{article}
\usepackage{epsfig}
\usepackage{amssymb}
\usepackage{graphicx}
\usepackage{latexsym}
\usepackage{bbold}
\usepackage{bbm}
\usepackage[mathscr]{eucal}
\usepackage[english]{babel}
\def\a{\alpha}
\def\b{\beta}

\def\d{\delta}

\def\m{\mu}

\def\r{\rho}

\def\t{\tau}

\def\y{\eta}

\def\f{\varphi}

\def\half{\frac{1}{2}} 
\def\ra{\rightarrow}

\def\mtil{\widetilde{m}}
\usepackage{color}
\definecolor{rosso}{cmyk}{0,1,1,0.4}
\definecolor{rossos}{cmyk}{0,1,1,0.55}
\definecolor{rossoc}{cmyk}{0,1,1,0.2}
\definecolor{blu}{cmyk}{1,1,0,0.3}
\definecolor{blus}{cmyk}{1,1,0,0.6}
\definecolor{bluc}{cmyk}{1,1,0,0.1}
\definecolor{verde}{cmyk}{0.92,0,0.59,0.25}
\definecolor{verdec}{cmyk}{0.92,0,0.59,0.15}
\definecolor{verdes}{cmyk}{0.92,0,0.59,0.4}

\def\circa#1{\,\raise.3ex\hbox{$#1$\kern-.75em\lower1ex\hbox{$\sim$}}\,}

\newcommand{\eV}{\,{\rm eV}}

\newcommand{\GeV}{\,{\rm GeV}}

\makeatletter
\newcommand{\beq}{\begin{equation}}
\newcommand{\eeq}{\end{equation}}
\newcommand{\bea}{\begin{eqnarray}}
\newcommand{\eea}{\end{eqnarray}}
\newcommand{\ba}{\begin{array}}
\newcommand{\ea}{\end{array}}
\newcommand{\bn}{\begin{enumerate}}
\newcommand{\en}{\end{enumerate}}
\newcommand{\bc}{\begin{center}}
\newcommand{\ec}{\end{center}}

\newcommand{\eps}{\epsilon}

\newcommand{\gsim}{\lower.7ex\hbox{$\;\stackrel{\textstyle>}{\sim}\;$}}
\newcommand{\lsim}{\lower.7ex\hbox{$\;\stackrel{\textstyle<}{\sim}\;$}}
\newcommand{\baz}{\begin{array}{cc}}
\newcommand{\bad}{\begin{array}{ccc}}
\def\gtap{\mathrel{ \rlap{\raise 0.511ex \hbox{$>$}}{\lower 0.511ex
   \hbox{$\sim$}}}} 
\def\ltap{\mathrel{ \rlap{\raise 0.511ex
   \hbox{$<$}}{\lower 0.511ex \hbox{$\sim$}}}}
   \newcommand{\deltaatm}{\mbox{$\Delta m^2_{\mathrm{A}}$}}
   \newcommand{\deltasol}{\mbox{$ \Delta m^2_{\odot}$}}
\newcommand{\dmsol}{\mbox{$\Delta m^2_{\odot}$}}
\newcommand{\dma}{\mbox{$\Delta m^2_{\rm A}$}}
   
   \newcommand{\betabeta}{\mbox{$(\beta \beta)_{0 \nu}$}}

\newcommand{\pmns}{\mbox{$ U$}}

\renewcommand{\thefootnote}{\fnsymbol{footnote}}

\begin{document}
\begin{titlepage}
\hfill
  \begin{center}

{\large \bf The Interplay Between the ``Low'' and ``High'' 
Energy CP-Violation in Leptogenesis\\[0.3cm]
}
\vspace*{7mm}

{\bf E.~Molinaro$^{a)}$
\footnote[1]{E-mail: molinaro@sissa.it} and 
S.~T.~Petcov$^{a,b)}$
\footnote[2]{Also at: Institute
of Nuclear Research and Nuclear Energy, Bulgarian Academy 
of Sciences, 1784 Sofia, Bulgaria.}}
 
\vskip 6pt

{\it $^{a)}$SISSA and INFN-Sezione di Trieste, Trieste I-34014, Italy}\\
{\it $^{b)}$IPMU, University of Tokyo, Tokyo, Japan}\\
\vspace{0.6cm}
\end{center}
\begin{abstract}
We analyse within the ``flavoured''
leptogenesis scenario of baryon asymmetry
generation, the interplay
of the ``low energy'' CP-violation,
originating from the
PMNS neutrino mixing matrix $U$,
and the ``high energy'' CP-violation
which can be present in the matrix of
neutrino Yukawa couplings,
$\lambda$, and can manifest itself
only in ``high'' energy scale processes.
The type I see-saw model with three
heavy right-handed Majorana neutrinos
having hierarchical spectrum is
considered.  The ``orthogonal'' parametrisation
of the matrix of neutrino Yukawa couplings,
which involves a complex orthogonal matrix
$R$, is employed. In this approach the
matrix $R$ is the source of ``high energy''
CP-violation. Results for normal hierarchical (NH)
and inverted hierarchical (IH)
light neutrino mass spectrum
are derived in the case of decoupling of the
heaviest RH Majorana neutrino. It is shown that
taking into account the contribution to
$Y_B$ due to the CP-violating phases in the
neutrino mixing matrix $U$
can change drastically the predictions for
$Y_B$, obtained assuming only ``high energy''
CP-violation from the $R$-matrix
is operative in leptogenesis.
In the case of IH spectrum, in particular,
there exist significant regions in the corresponding
parameter space where the purely ``high energy'' contribution
in $Y_B$ plays a subdominant role in the production of baryon
asymmetry compatible with the observations.
\end{abstract}
\end{titlepage}

\renewcommand{\thefootnote}{\arabic{footnote}}
\setcounter{footnote}{0}
\setcounter{page}{1}

\newpage

\section{Introduction}
\indent  
In the present article we 
investigate further the possible
connection between leptogenesis~\cite{FY,kuzmin} 
(see also, e.g.~\cite{LG1,LG2}) and the low 
energy CP-violation in the lepton
(neutrino) sector (for earlier discussions see,
e.g.~\cite{others,others2,PPR03,PRST05} and 
the references quoted therein).  
It was shown recently in \cite{PPRio106} 
that the CP-violation necessary for 
the generation of the observed baryon 
asymmetry of the Universe in the thermal 
leptogenesis scenario can be due exclusively
to the Dirac and/or Majorana CP-violating 
phases in the Pontecorvo-Maki-Nakagawa-Sakata (PMNS) 
neutrino mixing 
matrix~\cite{BPont57}, and thus can be directly 
related to the low energy CP-violation in the 
lepton sector, e.g. in neutrino oscillations, etc. 
(see also \cite{Branco06}). 
The baryon asymmetry is produced in the regime when 
the lepton flavour effects in leptogenesis 
~\cite{Barbieri99,Nielsen02,davidsonetal,davidsonetal2}
are significant (``flavoured'' leptogenesis). 
As was realised in \cite{davidsonetal,davidsonetal2},
the lepton flavour effects can play 
very important role in the
leptogenesis mechanism of 
baryon asymmetry generation.

  In ref. \cite{MPST07}, which was 
stimulated by some of the results 
obtained in \cite{PPRio106}, 
the dependence of the baryon 
asymmetry produced in ``flavoured'' leptogenesis 
on the lightest neutrino mass, 
${\rm min}(m_j)$, $j=1,2,3$,
when the requisite CP-violation 
is provided entirely by the PMNS matrix 
$U_{\rm PMNS}$,
was studied.
In certain rather general and physically 
interesting cases the generated baryon asymmetry
was found to depend strongly on the value of  
${\rm min}(m_j)$. For specific values of 
${\rm min}(m_j)$, in particular,
the asymmetry can be strongly enhanced 
(by a factor of $\sim 100$ or more) 
with respect to that predicted 
in the case of ${\rm min}(m_j) =0$.
This enhancement can make the predicted $Y_B$ 
compatible with the observations even when 
this is not the case for ${\rm min}(m_j) \cong 0$.
Some aspects of the matter-antimatter asymmetry 
generation in the ``flavoured'' leptogenesis
scenario in the case when the relevant CP-violation 
is due to the Majorana or Dirac CP-violating 
phases in $U_{\rm PMNS}$, were investigated also in 
\cite{SBDibari06,DiBGRaff06}.

 Here we analyse in detail the interplay
in ``flavoured'' leptogenesis 
of the ``low energy'' CP-violation, 
originating from the PMNS
neutrino mixing matrix,
and the ``high energy'' CP-violation
which can be present in the matrix of 
neutrino Yukawa couplings, 
$\lambda$, and can manifest itself 
only at some ``high'' energy scale.
Both types of CP-violation can, 
in general, have their origin in the 
neutrino Yukawa couplings.
The latter, as is well-known, 
are one of the basic ingredients of
the see-saw mechanism of neutrino mass 
generation~\cite{seesaw}, 
on which the leptogenesis theory is based. 
A widely recognised appealing features of the 
see-saw model include 
i) a natural explanation of the 
smallness of neutrino masses 
(see, e.g.~\cite{STPNu04,MoscowH3Mainz,Hann06}),
and that 
ii) it allows to relate, 
through the leptogenesis theory, 
the generation and the smallness of neutrino masses 
with the generation of the baryon
(matter-antimatter) asymmetry of the Universe, $Y_B$. 

   The minimal scheme in which leptogenesis can be realised 
is the non-supersymmetric version of the type I see-saw model 
with three (or two) heavy right-handed (RH) Majorana neutrinos, 
$N_j$, having masses $M_j$. 
The matrix of neutrino Yukawa couplings, 
$\lambda$, together with the Majorana mass 
matrix of the RH neutrinos, $M_R$, and the 
matrix of charged lepton Yukawa couplings, 
$\lambda^{lep}$,
plays a crucial role both in the see-saw 
mechanism and in leptogenesis.
In the basis in which 
$M_R$ and $\lambda^{lep}$ are diagonal
\footnote{This basis can be chosen 
without loss of generality and we will use
it in our further analysis.}, 
the matrix of neutrino Yukawa 
couplings $\lambda$ is the only source  
of CP-violation in the lepton sector.
Among the several possible parametrisations 
of $\lambda$ (see, e.g. \cite{PPR03}), the 
orthogonal one, involving a complex 
orthogonal matrix $R$ \cite{Casas01}, 
allows to relate in a rather direct manner 
the matrix $\lambda$ with the neutrino 
mixing matrix $U_{\rm PMNS} \equiv U$: 
$\lambda = (1/v)\sqrt{M} \, R\, \sqrt{m} \, U^{\dagger}$, 
where $M$ and $m$ are diagonal matrices formed by
the masses $M_j > 0$ and $m_k \geq 0$ of $N_j$ 
and of the light Majorana neutrinos $\nu_k$, $j,k=1,2,3$, 
and $v$ is the vacuum expectation
value of the Higgs doublet field. 
This parametrisation proved rather convenient in 
the analysis \cite{PPRio106}
of the possibility that the CP-violation
necessary for a successful leptogenesis 
could be provided by the 
Majorana and/or Dirac physical 
phases in the neutrino mixing matrix 
$U$. It permits to  investigate also the 
combined effect of the 
CP-violation due to the neutrino 
mixing matrix $U$ 
and the CP-violation due to the matrix $R$ 
in the generation of the baryon asymmetry in 
``flavoured'' leptogenesis. We will use the terms 
``low energy'' and ``high energy'' 
for the CP-violation originating 
respectively from the matrices $U$ and $R$.
The PMNS matrix $U$ is present 
in the weak charged lepton current 
and can be a source of CP-violation 
in, e.g. neutrino oscillations at 
``low'' energies
\footnote{As is well-known, 
only the Dirac phase in $U$ can be a source of 
CP-violation in neutrino oscillations;
the probabilities of oscillations of
flavour neutrinos do not depend on the 
Majorana phases in $U$ \cite{BHP80,Lang87}.}
$E \sim M_Z$ 
(see, e.g. \cite{BiPet87,PKSP3nu88,Future}). 
The matrix $R$, as is 
well-known, does not affect the ``low'' 
energy neutrino mixing phenomenology.
The two matrices $U$ and $R$
are, in general, independent. 
It should be noted, however, 
that in certain specific cases 
(of, e.g. symmetries and/or texture zeros)
of the matrix $\lambda$ of neutrino Yukawa couplings, 
there can exist a relation 
between (some of) the CP-violating phases in 
$U$ and (some of) the CP-violating 
parameters in $R$ (see, e.g. \cite{PRST05,PSMoriond06}).

  The source of the requisite CP-violation 
in ``flavoured'' leptogenesis can, in principle, be    
the matrix $R$, the PMNS matrix $U$, 
or both the matrices $R$ and $U$. 
The division between CP-violation 
due to the PMNS matrix $U$ and that due to the 
matrix $R$ at ``high'' energies, 
e.g. in leptogenesis, needs some clarification.
If the matrix $R$ satisfies the general 
CP-invariance constraints (having real or purely 
imaginary elements \cite{PPRio106}),
while the PMNS matrix, and correspondingly, the 
matrix of neutrino Yukawa couplings $\lambda$ do 
not satisfy these constraints, 
we will consider the CP-violation as originating 
from the neutrino mixing matrix $U$, i.e. 
from the Dirac and/or Majorana phases in $U$.
If, however, the Dirac and Majorana phases in $U$ 
take CP conserving values, while the matrix $R$, 
and the Yukawa couplings $\lambda$ 
do not satisfy the constraints following from 
the requirement of CP-invariance, 
the CP-violation will manifest itself 
only in ``high'' energy phenomena 
(like, e.g. leptogenesis)
and will be due to the matrix $R$. 
In this case there will be
no (observable) effects of violation 
of CP symmetry at ``low'' energies in 
phenomena caused by the neutrino mixing 
(neutrino oscillations, neutrinoless double 
beta (\betabeta-) decay \cite{BPP1,STPFocusNu04}, etc.).
When neither $U$ nor $R$ satisfy 
the CP-invariance conditions,
both $U$ and $R$ will be 
sources of CP-violation effects 
at ``high'' energies .

    In the present article we 
investigate the possible interplay  
between the ``low energy'' CP-violation 
due to the Dirac and/or Majorana CP-violating 
phases in the PMNS matrix $U$,
and the ``high energy'' CP-violation 
originating from the matrix $R$, 
in ``flavoured'' leptogenesis 
We work within the simplest type 
I see-saw scenario with three heavy RH Majorana 
neutrinos $N_j$, $j=1,2,3$.  The latter are
assumed to have a hierarchical mass spectrum, $M_1 \ll M_{2,3}$.
In what concerns the light neutrino masses,
we consider two types of spectrum allowed 
by the existing data (see, e.g. \cite{STPNu04}), namely,
the normal hierarchical (NH), $m_1 \ll m_2 < m_3$, and 
the inverted hierarchical (IH), $m_3 \ll m_1 < m_2$,
and present detailed results for these two spectra. 

   There is practically no overlap between
the results obtained in our work
and those found in  ref. \cite{davidsonetal2}.
One of the objectives of the analysis 
in \cite{davidsonetal2} was to find out 
whether there can be large differences between the
baryon asymmetry of the Universe, 
predicted in the flavoured leptogenesis
scenario (the two- or three- flavour regime),
and the asymmetry derived in unflavoured
leptogenesis (the one-flavour regime).
The authors of \cite{davidsonetal2} were not interested in
(and did not try to provide an answer to) the question of the
the relative magnitude of, and the interplay between, the contributions to
the baryon asymmetry due to the ``low-energy'' CP violating phases in the
PMNS matrix and that due to the ``high energy'' CP violating phases in the
$R$ matrix, which is the main subject of investigation in our work.

    Negative results regarding the possible interplay
between the ``high energy'' and ``low energy''
CP violation in ``flavoured'' leptogenesis with
hierarchical heavy (RH) and light Majorana neutrinos
were reported in \cite{SDavid07}. 
The results reported in \cite{SDavid07} were obtained
for i) ${\rm min}(m_j) =0$ and specific
texture zero in the $R$-matrix,
or  ii) for a value of the mass of the
lightest RH neutrino
\footnote{Private communication by S. Davidson. 
We thank S. Davidson for clarifications regarding the analysis 
performed in \cite{SDavid07}.}
$M_1 = 10^{10}$ GeV.
In case ii) the lightest neutrino mass
${\rm min}(m_j)$ was allowed to vary within
the interval $0 \leq {\rm min}(m_j) \leq 10^{-3}$ eV.
However, in both these cases the contribution 
in $Y_B$ due to the ``low energy'' CP violation
is strongly suppressed. The contribution under discussion can
be relevant for the production of $Y_B$ compatible with
the observations provided \cite{PPRio106}
$M_1 \gtap 4\times 10^{10}$ GeV.  In case i) it can be
relevant if one considers values of
$M_1 \gtap 4\times 10^{10}$ GeV and of
${\rm min}(m_j) \gtap 5\times 10^{-4}$ eV \cite{MPST07}.  

  The lepton flavour effects can be significant in leptogenesis
in the case of hierarchical spectrum of the heavy Majorana neutrinos, 
provided the mass of the lightest one $M_1$ satisfies the constraint
\cite{davidsonetal,davidsonetal2} (see also \cite{DiBGRaff06}): 
$M_1 \ltap 10^{12}~{\rm GeV}$.  In this case the predicted value of the
baryon asymmetry depends explicitly (i.e.  directly) on $U$ and on the
CP-violating phases in $U$. Using this fact it was shown in 
\cite{PPRio106} that the observed baryon asymmetry $Y_B$ can be 
produced even if the only source of CP-violation in leptogenesis 
is the Majorana  and/or Dirac phase(s) in the PMNS matrix 
\footnote{The same conclusion was shown to be 
valid also for quasi-degenerate in mass 
heavy 
Majorana neutrinos \cite{PPRio106}.}  
$U$. These results were demonstrated to hold 
both for normal hierarchical (NH) and inverted
hierarchical (IH) spectrum of masses of the light 
Majorana neutrinos.
They were obtained for CP-conserving elements 
of the orthogonal matrix $R$.  
The CP-invariance constraints imply 
\cite{PPRio106} that the matrix $R$ could 
conserve the CP-symmetry if its elements are real
or purely imaginary.

   Our analysis is performed for negligible 
renormalisation group (RG) running of $m_j$
and of the parameters in the PMNS matrix $\pmns$ from
$M_Z$ to $M_1$ (see, e.g. \cite{rad1,rad2,PST06})~
\footnote{We have checked by explicit calculations 
using the equations describing the RG running 
of the neutrino mixing parameters
given in http://www.ph.tum.de/\~mratz/AnlyticFormulae/ 
(see \cite{rad2}), 
that the running of the neutrino mixing angles, 
the Majorana and Dirac phases 
in the neutrino mixing matrix and of 
the neutrino masses, is negligible in both cases 
of NH and IH light neutrino mass spectrum
studied. We have verified,
in particular, that for the values
of $\sin\theta_{13}$ of interest 
for our discussion, 
the running of the Dirac phase
in the case of NH spectrum is so 
small that can be safely neglected.
The running of the Dirac phase 
in the case of IH spectrum 
with negligible lightest neutrino mass 
$m_3$, as is well-known, is negligible.}.
Throughout the present work we use the 
standard parametrisation of the PMNS matrix:
\bea 
\label{eq:Upara}
\pmns = \left( \bad 
c_{12} c_{13} & s_{12} c_{13} & s_{13}e^{-i \delta}  \\[0.2cm] 
 -s_{12} c_{23} - c_{12} s_{23} s_{13} e^{i \delta} 
 & c_{12} c_{23} - s_{12} s_{23} s_{13} e^{i \delta} 
& s_{23} c_{13}  \\[0.2cm] 
 s_{12} s_{23} - c_{12} c_{23} s_{13} e^{i \delta} & 
 - c_{12} s_{23} - s_{12} c_{23} s_{13} e^{i \delta} 
 & c_{23} c_{13} \\ 
                \ea   \right) 
~{\rm diag}(1, e^{i \frac{\alpha_{21}}{2}}, e^{i \frac{\alpha_{31}}{2}})
\eea\newline\newline
%

\noindent where $c_{ij} \equiv \cos\theta_{ij}$, $s_{ij} \equiv
\sin\theta_{ij}$, $\theta_{ij} = [0,\pi/2]$, $\delta = [0,2\pi]$ is
the Dirac CP-violating (CPV) phase and $\alpha_{21}$ and $\alpha_{31}$
are the two Majorana CPV phases \cite{BHP80,SchValle80Doi81},
$\alpha_{21,31} = [0,4\pi]$.  All our numerical results 
are obtained for the 
best fit values of the solar and atmospheric neutrino
oscillation parameters
\cite{BCGPRKL2,TSchwSNOW06,Fogli06},
$\deltasol$, $\sin^2 \theta_{12}$ and $\deltaatm$, 
$\sin^22\theta_{23}$:
\begin{eqnarray}
\label{deltaatmvalues}
\deltasol = \Delta m^2_{21} =
 8.0 \times 10^{-5} \eV^2,~~\sin^2 \theta_{12} = 0.30,\\ [0.25cm]
|\deltaatm| = |\Delta m^2_{31(32)}| = 2.5 \times 10^{-3} \eV^2,~~
\sin^2 2\theta_{23} = 1.
\end{eqnarray}
%
In certain cases
\footnote{Using the latest data 
from the KamLAND and SNO experiments  
in the global  neutrino oscillation analysis 
one obtains somewhat different 
best fit values of $\deltasol$,  
and $|\deltaatm|$ \cite{TSchw08}:
$\deltasol = 7.65 \times 10^{-5}~{\rm eV^2}$, 
$|\deltaatm| = 2.4 \times 10^{-3}~{\rm eV^2}$.
The results of our analysis do not 
change if we use these best fit values.}
the predictions for $|Y_B|$ 
are very sensitive to the variations of 
$\sin^22\theta_{23}$ within  
its 95\% C.L.  allowed range:
\beq
0.36 \ltap \sin^2 \theta_{23} \ltap 0.64,~95\%~{\rm C.L.}
\label{th122395}
\eeq
%
We also use the upper limit on the CHOOZ mixing angle
$\theta_{13}$ \cite{CHOOZ,BCGPRKL2,TSchw08}:
\beq
\sin^2\theta_{13} < 0.035~(0.056)\,,~~~
95\%~(99.73\%)~{\rm C.L.}\,.
\label{th13}
\eeq
%

%
%
\section{Baryon Asymmetry from ``Low'' and ``High'' 
Energy CP-Violation
}
%
%
\indent Following \cite{PPRio106,MPST07}, we perform the analysis in the
framework of the simplest type I see-saw scenario.  It includes the
Lagrangian of the Standard Model (SM) with the addition of three heavy
right-handed Majorana neutrinos $N_{j}$ ($j=1,2,3$) with masses $0 <
M_{1} < M_{2} < M_{3}$ and Yukawa couplings $\lambda_{j l}$,
$l=e,\mu,\tau$.  We will work in the basis in which i) the Yukawa
couplings for the charged leptons are flavour-diagonal, and ii) the
Majorana mass term of the RH neutrino fields is also diagonal.  The
heavy Majorana neutrinos are assumed to possess a hierarchical mass
spectrum, $M_1 \ll M_2 \ll M_3$.

In what follows we will use the well-known ``orthogonal
parametrisation`` of the matrix of neutrino Yukawa couplings
\cite{Casas01}:
\begin{equation}
\label{R}
\lambda = \frac{1}{v} \,  \sqrt{M} \, R\, \sqrt{m}\, U^{\dagger}\;,
\end{equation} 
%
\noindent where $R$ is, in general, a complex orthogonal matrix,
$R~R^T = R^T~R = {\bf 1}$, $M$ and $m$ are diagonal matrices formed by
the masses of $N_j$ and of the light Majorana neutrinos $\nu_j$, $M
\equiv {\rm Diag}(M_1,M_2,M_3)$, $m \equiv {\rm Diag}(m_1,m_2,m_3)$,
$M_j > 0$, $m_k \geq 0$, and $v = 174$ GeV is the vacuum expectation
value of the Higgs doublet field. In contrast to \cite{PPRio106,MPST07},
we shall assume that the matrix $R$ is not CP-conserving and thus 
has complex elements.

  In the case of ``hierarchical'' heavy Majorana neutrinos $N_j$, the
CP-violating asymmetries, relevant for leptogenesis, are generated in
out-of-equilibrium decays of the lightest one 
\footnote{It is well known that,
in thermal leptogenesis with hierarchical 
heavy Majorana neutrinos, the contribution
of the next to lightest one, $N_2$, 
to the baryon asymmetry can be significant \cite{Engelhard:2006yg}.
We limited our analysis to the conventional $N_1$ dominated scenario 
because we didn't want to assume particular constraints on the neutrino
Yukawa couplings, such that, for example, can imply no
$N_1$ washout effects during $N_2$ leptogenesis. 
}, $N_1$. The asymmetry
in the lepton flavour $l$ (lepton charge $L_l$) is given by
\cite{davidsonetal,davidsonetal2}:
\begin{eqnarray}
\label{epsa1}
\epsilon_{l}& \cong & -\frac{3 M_1}{16\pi v^2} \frac{{\rm Im}\left(
\sum_{j k} 
m_j^{1/2}m_k^{3/2} 
U^*_{lj}U_{lk}
R_{1j }R_{1k}\right)}{\sum_i m_i \left|R_{1i}\right|^2}\,.
\end{eqnarray}
%
Obviously, the total asymmetry 
($\epsilon_{e} + \epsilon_{\mu} + \epsilon_{\tau}$)
does not depend on the neutrino mixing matrix $U$ 
(E. Nardi et al., ref. \cite{davidsonetal}).

 We shall assume that the baryon asymmetry is produced 
in the ``two-flavour'' regime in leptogenesis 
\cite{davidsonetal,davidsonetal2}.
This regime is realised at temperatures 
$10^{9}~{\rm GeV} \ltap T \sim M_1 \ltap 10^{12}~{\rm GeV}$.
For $T \sim M_1$ from the indicated interval, 
the Boltzmann evolution 
of the asymmetry $\epsilon_{\tau}$ in the
$\tau-$flavour (lepton charge $L_\tau$ 
of the Universe) is distinguishable from 
the evolution of the $(e + \mu)-$flavour 
asymmetry $\epsilon_{e}+\epsilon_{\mu}$. 
In the two-flavour regime, 
the baryon asymmetry 
\footnote{The expression we give is
of the baryon asymmetry normalised to 
the entropy density, see, e.g.  \cite{PPRio106}.} 
predicted in the case of interest is given by
\cite{davidsonetal2}: 
\begin{eqnarray}
Y_B &\!\!\cong\!\!&  -\frac{12}{37 g_*}\left(\epsilon_2\, 
\eta\left(\frac{417}{589}\,\widetilde{m}_2\right) 
+ \epsilon_\tau\,\eta\left(\frac{390}{589}
\,\widetilde{m}_\tau\right)\right)\nonumber \,. 
\label{YB2f}
\end{eqnarray}
%
Here $g_* = 217/2$ is the number of relativistic
degrees of freedom, $\epsilon_2 = \epsilon_{e} + \epsilon_{\mu}$,
$\widetilde{m}_2=\widetilde{m}_e+ \widetilde{m}_\mu$, 
$\widetilde{m}_l$, $l=e,\mu,\tau$, is the 
``wash-out mass parameter'' for the asymmetry
in the lepton flavour $l$ \cite{davidsonetal,davidsonetal2},
\begin{eqnarray}
  \widetilde{m}_l 
  &=& \left|\sum_{k}R_{1k}\, m_k^{1/2}\, U_{lk}^*\right|^2
  \,, ~~~l=e,\mu,\tau\,,
\label{tildmal1}
\end{eqnarray}
%
and $\eta(390\widetilde{m}_\tau/589)\cong
\eta(0.66\widetilde{m}_\tau)$ and $\eta(417\widetilde{m}_2/589) \cong
\eta(0.71\widetilde{m}_2)$ are the efficiency factors for generation
of the asymmetries $\epsilon_{\tau}$ and $\epsilon_2$. The efficiency
factors are well approximated by the expression \cite{davidsonetal2}:
\begin{equation}
\eta\left(X\right) \cong 
 \left(
\frac{8.25\times 10^{-3}{\rm eV}}{X}
 + \left(
\frac{X}{2\times 10^{-4}{\rm eV}} 
\right)^{1.16}
\right)^{-1}.
\label{eta1}
\end{equation}
%
\newline
 For the observed value of the baryon asymmetry $Y^{obs}_B$, to be 
reproduced by leptogenesis, we will use in our analysis
$\bar{Y}^{obs}_B = 8.6\times 10^{-11}$ and the ``conservative'' 
interval $8.0\times 10^{-11} \leq Y^{obs}_B \leq 9.2\times 10^{-11}$.

 For complex $R_{1j}$ of interest
it proves convenient to cast the asymmetries 
$\epsilon_{l}$, $l=e,\mu,\tau$, in the form:
\newline
\begin{eqnarray}
\eps_\ell & = & -\frac{3\,M_1}{16\,\pi\, v^2}\,
\frac{1}{\sum_k \,m_k |R_{1k}|^2}\,
\left\{\sum_\b\,m_\b^2\,|R_{1\b}|^2\,|U_{\ell\b}|^2\,
\sin 2\tilde{\f}_{1\b}\,
+\, \sum_{\b}\,\sum_{\r>\b}\,\sqrt{m_\b\,m_\r}\,|R_{1\b}R_{1\r}|\right.
\nonumber\\
\label{CP-asym-b}\\
& \times & \left.\left[(m_\r\,-\,m_\b)\,\cos(\f_{\b\r})\,
{\rm Im}\left(U_{\ell\b}^*\,U_{\ell\r}\right)\,
+\,(m_\r+m_\b)\,\sin(\f_{\b\r})\,
{\rm Re}\left(U_{\ell\b}^*\,U_{\ell\r}\right)\right]\frac{}{×}\right\}\,,
\nonumber
\end{eqnarray}
%
where we have used
\begin{equation}
R_{1j}\,\equiv\,|R_{1j}|\,e^{i\,\tilde{\f}_{1j}}\,,
\;\;\;\;\;\;\;\f_{ij}\,\equiv\,\tilde{\f}_{1i}\,
+\,\tilde{\f}_{1j}\,.\;\;\;\;\;\;\;
\label{Rph1}
\end{equation}
%
The first term in the curly brackets in 
eq. (\ref{CP-asym-b}) represents the contribution
to $\eps_\ell$ from  the ``high energy'' CP-violation, 
originating entirely from the matrix $R$, 
while the terms in the square brackets 
are ``mixed'', i.e. they are due both 
to the ``low'' and ``high'' energy 
CP-violation, generated by the 
neutrino mixing matrix $U$ and 
by the matrix $R$. In what follows we will call the 
CP-violating phases $\tilde{\f}_{1j}$ 
associated with the matrix $R$, ``$R$-phases'' or 
``high energy'' phases.
Obviously, if $\tilde{\f}_{1j} = k_j\pi/2$, $k_j=0,1,2,...$, 
$j=1,2,3$, the ``high energy'' term is zero, while
the ``mixed'' term reduces to a ``low energy'' 
term in the sense that, with exception of very 
special cases (see \cite{PPRio106}),
the only source of CP-violation 
in leptogenesis will be the PMNS matrix $U$.
The expression (\ref{CP-asym-b}) for 
$\eps_\ell$ implies that,
since $R_{1j}$ satisfy the orthogonality 
condition $R_{11}^2 + R_{12}^2 + R_{13}^2 = 1$,
we can have $\eps_\ell \neq 0$  
only if at least two of the three elements 
$R_{1j}$ of the first row of $R$  
are different from zero. We will be interested 
primarily in the effect the
``high energy'' and the ``mixed'' terms 
have on the predicted value of $Y_B$.
 
 
\section{Normal Hierarchical 
Neutrino Mass Spectrum}


 In the case of NH light neutrino mass spectrum we have 
$m_1 \ll m_2 < m_3$, and consequently, 
$m_2 \cong \sqrt{\dmsol} \cong 9\times 10^{-3}$ eV, 
$ m_3 \cong \sqrt{\dma} \cong 5\times 10^{-2}$ eV.
We shall investigate here the case
of negligibly small lightest neutrino mass $m_1$.
More precisely, we will assume 
that the terms proportional to 
$(m_1)^n$, $n=1/2;~1;~2$, in the expressions for 
$\eps_{\tau}$, $\eps_{2}$,
$\widetilde{m}_\tau$ and $\widetilde{m}_2$
are significantly smaller than the 
terms proportional to $(m_{2,3})^n$, 
and can be neglected.
In what follows we set $m_1 = 0$ for simplicity.
The asymmetry $\eps_{\tau}$ in this case 
takes the form: 
\begin{eqnarray}
\label{epstau}
\eps_\t & \cong & -\frac{3\,M_1}{16\,\pi\, v^2}\,\frac{\sqrt{\dma}}
{\left(\frac{\dmsol}{\dma}\right)^{1/2}\,|R_{12}|^2\,+\,|R_{13}|^2}
\left\{\left(\frac{\dmsol}{\dma}\right)\,|R_{12}|^2\,|U_{\t 2}|^2\,
\sin2\tilde{\f}_{12}
\right.\nonumber\\\nonumber\\ 
& + & \left.|R_{13}|^2\,|U_{\t3}|^2\,
\sin2\tilde{\f}_{13}\,
+\,\left(\frac{\dmsol}{\dma}\right)^{1/4}\,|R_{12}|\,|R_{13}|\,
\left[\left(1-\frac{\sqrt{\dmsol}}{\sqrt{\dma}}\right)
\cos(\tilde{\f}_{12} +\tilde{\f}_{13})\,
{\rm Im}\left(U_{\t2}^*\,U_{\t3}\right)
\right.\right.\nonumber\\
& + & \left.\left. 
\left(1+\frac{\sqrt{\dmsol}}{\sqrt{\dma}}\right)
\sin(\tilde{\f}_{12} +\tilde{\f}_{13})\,
{\rm Re}\left(U_{\t2}^*\,U_{\t3}\right)\right]\right\}\,.
\end{eqnarray}
%
\noindent It is easy to show, 
taking into account the unitarity of the matrix $U$, 
that the expression for the 
CP-asymmetry $\eps_{2}$ 
can be simply obtained from the expression 
for $\eps_\tau$ :
\begin{equation}
\eps_2\,\equiv\,\eps_e\,+\eps_\m\,
=\,\eps_\t(|U_{\t k}|^2\ra 1-|U_{\t k}|^2\,,\, 
U_{\t 2}^*U_{\t 3}\ra -U_{\t 2}^*U_{\t 3})\,,~~k=2,3\,.
\label{eps2}
\end{equation}
%
\indent The first term in the brackets in eq. (\ref{epstau}) 
is suppressed by the factor $\dmsol/\dma \cong 0.03$.
A more detailed study shows that 
it always plays a subdominant role 
in the generation of baryon asymmetry 
compatible with observations 
and can be safely neglected.

  It should be clear from the expressions (\ref{epstau}) 
and (\ref{eps2}) for the asymmetries 
$\eps_{\tau}$ and $\eps_{2}$ that the CP-violation 
due to the PMNS matrix $U$ can play a significant 
role in leptogenesis only if 
the ``mixed'' term proportional to 
$|R_{12}R_{13}|$ in eq. (\ref{epstau}) 
is comparable in magnitude, 
or exceeds, the ``high energy'' term 
$\propto |R_{13}|^2|U_{\t3}|^2\sin2\tilde{\f}_{13}$. 
The latter will not give a contribution 
to the asymmetries $\eps_{\tau}$ and $\eps_{2}$
if $\sin2\tilde{\f}_{13} = 0$, i.e. 
if $R_{13}$ is real or purely imaginary 
\cite{PPRio106}.

  The elements of the matrix $R$ are constrained by the 
orthogonality condition: $R_{11}^2 + R_{12}^2 + R_{13}^2 = 1$. 
In the case of ``small'' lightest neutrino mass $m_1$ 
under consideration, the $R_{11}$ element does
not appear in the expressions for 
$\eps_{\tau}$, $\eps_{2}$,
$\widetilde{m}_\tau$ and $\widetilde{m}_2$, 
which are relevant for the calculation of the 
baryon asymmetry $Y_B$. We will analyse in what 
follows the possibility of relatively small $|R_{11}|$, 
so that the term $R_{11}^2$ in the orthogonality 
condition can be neglected. This is realised if
$|R_{11}|^2 \ll {\rm min}(1,|R_{12}|^2|\sin2\tilde{\f}_{12}|)$.
Under this condition we can set $R_{11} = 0$ 
in all our further considerations.
Let us note that we get $R_{11} = 0$, 
e.g. in the case of decoupling of the heaviest RH
Majorana neutrino $N_3$ \cite{PRST05,IR041}, 
leading effectively to the so-called 
``$3\times 2$'' see-saw model 
\footnote{In the context of ``flavoured'' 
leptogenesis the case of $N_3$ decoupling 
was discussed earlier in 
\cite{davidsonetal2,PPRio106}. 
However, our analysis 
practically does not overlap
with the analyses performed in 
\cite{davidsonetal2,PPRio106}. 
Let us note also that 
in the SUSY version of the 
``$3\times 2$'' see-saw model 
it  can be experimentally feasible, 
in principle, to reconstruct 
the $R$-matrix, as discussed in  \cite{AIb06}. 
If achieved, such a reconstruction
could shed light on the origin of 
the CP violation in leptogenesis.} 
\cite{FGY03}.

  For negligible $|R_{11}|^2$, 
the orthogonality condition for 
the elements of $R$ can be written
in terms of two equations involving 
the absolute values and 
the phases of $R_{12}$ and $R_{13}$:

\begin{equation}
|R_{12}|^2\,
\cos 2\tilde{\f}_{12}
+\,|R_{13}|^2\,
\cos 2\tilde{\f}_{13}\,=\,1\,,
\label{orth1}
\end{equation}
\begin{equation}
|R_{12}|^2\,
\sin 2\tilde{\f}_{12}
+\,|R_{13}|^2\,
\sin 2\tilde{\f}_{13}\,=\,0\,.
\label{orth2}
\end{equation}
%
Obviously, we should have 
${\rm sgn}(\sin 2\tilde{\f}_{12}) 
= - {\rm sgn}(\sin 2\tilde{\f}_{13})$.
Using these equations we can express 
the phases $\tilde{\f}_{12}$ and 
$\tilde{\f}_{13}$ in terms of 
$|R_{12}|^2$ and $|R_{13}|^2$:
\begin{eqnarray}
\cos 2\tilde{\f}_{12} & = &
\frac{ 1 + |R_{12}|^4 - |R_{13}|^4}{2|R_{12}|^2}\,,~~
\sin 2\tilde{\f}_{12}\,=\, \pm \sqrt{1 - \cos^2 2\tilde{\f}_{12}}\,,
\label{tf12} \\
\cos 2\tilde{\f}_{13} & = &
\frac{ 1 - |R_{12}|^4 + |R_{13}|^4}{2|R_{13}|^2}\,,~~
\sin 2\tilde{\f}_{13}\,=\, \mp \sqrt{1 - \cos^2 2\tilde{\f}_{13}}\,.
\label{tf13} 
\end{eqnarray}
%
The fact that $-1 \leq \cos 2\tilde{\f}_{12(13)}\leq 1$ 
leads to the following conditions:
\begin{eqnarray}
\large(1 + |R_{12}|^2\large)^2 \geq |R_{13}|^4\,,~~
\large(1 - |R_{12}|^2\large)^2 \leq |R_{13}|^4\,;
\label{tf12R} \\
\large(1 + |R_{13}|^2\large)^2 \geq |R_{12}|^4\,,~~
\large(1 - |R_{13}|^2\large)^2 \leq |R_{12}|^4\,.
\label{tf13R} 
\end{eqnarray}
%
Alternatively,
one can express 
$|R_{12}|^2$ and $|R_{13}|^2$ as functions 
of the phases:
\begin{eqnarray}
|R_{12}|^2 & = & 
\frac{\sin2\tilde{\f}_{13}}
{\sin2(\tilde{\f}_{13} - \tilde{\f}_{12})}\,,
\nonumber\\
\label{R-elements}\\ 
\nonumber
|R_{13}|^2 & = & 
-\, \frac{\sin2\tilde{\f}_{12}}
{\sin2(\tilde{\f}_{13} - \tilde{\f}_{12})}\,. 
\nonumber
\end{eqnarray}
%
\noindent We will consider values of 
the phases $\tilde{\f}_{12}$ and $\tilde{\f}_{13}$
from the interval [0,$2\pi$]. 
The positivity of $|R_{12}|^2$ and $|R_{13}|^2$
implies the following constraints on the
allowed ranges of
$\tilde{\f}_{12}$ and $\tilde{\f}_{13}$:
\begin{eqnarray}
\label{intervall1}
& k\pi \leq \, \tilde{\f}_{13}\,\leq \,(2k+1)\frac{\pi}{2}\,,~~~
\tilde{\f}_{13}\, - \frac{\pi}{2} -  k'\pi 
< \, \tilde{\f}_{12}\,\leq \,(k -k')\pi\,;\\ [0.25cm] 
& (2k + 1)\frac{\pi}{2} \leq \, \tilde{\f}_{13}\,\leq \,(k+1)\pi\,,~~~
(k -k')\pi\, \leq \, \tilde{\f}_{12}\, <
\tilde{\f}_{13}\, - \frac{\pi}{2} -  k'\pi\,,
\label{intervall2}
\end{eqnarray}
%
where $k=0,1,2,3$ and $k'=0,\pm 1,\pm 2, \pm 3$.
 
  We will be interested in the case when the ``mixed'' 
term $\propto |R_{12}R_{13}|$ in the expression (\ref{epstau})
for the CP-asymmetry $\eps_\t$ is sufficiently large 
and gives either a dominant contribution 
to $\eps_\t$, or at least a comparable 
one to that due to the ``high energy'' term 
$\propto |R_{13}|^2|U_{\t3}|^2\sin2\tilde{\f}_{13}$.
Accordingly, 
it is useful to know the values
$|R_{12}|$ and $|R_{13}|$ 
which maximise the function
\begin{equation}
F_1(|R_{12}|,|R_{13}|)\,=\,\frac{|R_{12}|\,|R_{13}|}
{\left(\frac{\dmsol}{\dma}\right)^{1/2}\,|R_{12}|^2\,
+\,|R_{13}|^2}\,.
\label{F_1}
\end{equation}
%
The maximum of $F_1(|R_{12}|,|R_{13}|)$
is obtained for \cite{PPRio106}
$|R_{12}|/|R_{13}|=(\dma/\dmsol)^{1/4}\cong 2.4$, and at 
the maximum we have $F_1^{{\rm max}} = 
0.5\,(\dma/\dmsol)^{1/4}\cong 1.2$.
At $|R_{12}|/|R_{13}|=(\dma/\dmsol)^{1/4}$,
the corresponding function 
in the ``high energy'' term in $\eps_\t$,
\begin{equation}
F_3(|R_{12}|,|R_{13}|)\,=\, 
\frac{|R_{13}|^2}{\left(\frac{\dmsol}{\dma}\right)^{1/2}
|R_{12}|^2+|R_{13}|^2}\,,
\label{F_3}
\end{equation}
%
takes the value 0.5, which is smaller 
only by a factor of 2 than its maximal 
possible value. The latter, however, 
takes place at $|R_{12}|=0$, for which 
$\eps_\t = \eps_2 = 0$. 

   The wash-out mass parameters in the case of interest 
are given by:
\begin{eqnarray}
\label{tmtau1}
\mtil_{\t} & = & \sqrt{\dmsol}\,|R_{12}|^2\,|U_{\t2}|^2\, 
+ \,\sqrt{\dma}\,|R_{13}|^2\,|U_{\t3}|^2\nonumber\\\\ 
& + & 2(\dmsol\dma)^{1/4}|R_{12}||R_{13}|
{\rm Re}\left(e^{
i(\tilde{\f}_{12}-\tilde{\f}_{13})}
U_{\t2}^*U_{\t3}\right)\nonumber
\\
\mtil_2 & = & \sqrt{\dmsol}\,|R_{12}|^2\,+\,\sqrt{\dma}\,|R_{13}|^2\, 
- \,\mtil_{\t}\,. 
\label{tm21}
\end{eqnarray}
%
Note that the 
``high energy'' phase difference 
$(\tilde{\f}_{12} - \tilde{\f}_{13})$ 
in the expressions for $\mtil_\t$ and $\mtil_2$ 
adds up to the Majorana phase (difference) 
$\alpha_{32}/2$, where 
$\alpha_{32} \equiv (\alpha_{31} - \alpha_{21})$.

  We will analyse next the combined effects of the 
``high energy'' and ``low energy'' CP-violating phases 
on the generation of the baryon asymmetry.

%
\subsection{CP Violation Due to 
Majorana Phase in $U_{\rm PMNS}$ 
and $R-$Phases}
%

\hspace{0.8cm}  Consider first the possibility that 
the baryon asymmetry $|Y_B|$ is generated 
by the combined effect of CP-violation due to
the Majorana phases in the PMNS matrix $U$
and the phases $\tilde{\f}_{12}$ and $\tilde{\f}_{13}$
of the orthogonal matrix $R$. 
The Dirac phase $\delta$ will be assumed to take 
a CP-conserving value: $\delta = \pi k$, 
$k=0,1,2,...$. The CP-violating asymmetries 
$\eps_\t$ and $\eps_2$
and the wash-out mass parameters $\mtil_\t$ and $\mtil_2$
depend in the case under study on 
the Majorana phase difference
$\a_{32}$. We will be interested in 
the interplay between the effects of 
the phases $\a_{32}$, $\tilde{\f}_{12}$ 
and $\tilde{\f}_{13}$
on the predicted value of $|Y_B|$.

  In the case under consideration the asymmetry $\eps_\t$
can be written in the form:
\begin{eqnarray}
\label{epstMaj1}
\eps_\t & \cong & -\, \frac{3\,M_1\,\sqrt{\dma}}{16\,\pi\,v^2}\,
\left\{\, F_3\, |U_{\t3}|^2\, 
\sin 2\tilde{\f}_{13} \right.
\nonumber\\
& - & \left. \left(\frac{\dmsol}{\dma}\right)^{\frac{1}{4}}\, F_1\, 
|U_{\t2}^*U_{\t3}|\, 
\left[\,\sin(\f_{23} + \frac{\a_{32}}{2}) + 
\left(\frac{\dmsol}{\dma}\right)^{\frac{1}{2}}\,  
\sin(\f_{23} - \frac{\a_{32}}{2})
\right]\right\}\,,
\end{eqnarray}
%
where $\f_{23}=\tilde{\f}_{12}+\tilde{\f}_{13}$,
the functions $F_1$ and $F_3$ are defined respectively 
by eqs. (\ref{F_1}) and (\ref{F_3}) and we have 
used the fact that for $\delta = \pi k$,
$(exp(-i\alpha_{32}/2)U_{\t2}^*U_{\t3}) = 
- (c_{12}s_{23} \pm s_{12}c_{23}s_{13})c_{23}c_{13} = 
-|U_{\t2}^*U_{\t3}|$.
The asymmetry $\eps_2$ 
can be obtained from eq. (\ref{epstMaj1})
by replacing $|U_{\t3}|^2$ with 
$(1 -|U_{\t3}|^2)$ and by changing the 
minus sign in front of the term $\propto F_1$ 
to plus sign. The expression for the baryon asymmetry 
has the form
\begin{eqnarray}
\label{YBMaj1}
Y_B & \cong & Y^0_B\, \left (\, A_{\rm HE} + A_{\rm MIX} \right )\,, 
\end{eqnarray}
%
where 
\begin{eqnarray}
Y^0_B \cong \frac{12}{37 g_*}\,\frac{3\,M_1\,\sqrt{\dma}}{16\,\pi\,v^2}
\cong 3\times 10^{-10}\,\left ( \frac{M_1}{10^9~{\rm GeV}}\right )\, 
\left (\frac{\sqrt{\dma}}{5\times 10^{-2}~{\rm eV}}\right )\,.
\label{Y0B}
\end{eqnarray}
%
\begin{eqnarray}
A_{\rm HE}\, = \, F_3\,  
\sin 2\tilde{\f}_{13} 
\left [~|U_{\t3}|^2\,\eta(0.66\widetilde{m}_\tau)
+ (1 - |U_{\t3}|^2)\,\eta(0.71\widetilde{m}_2)\,\right ]\,,
\label{AHE1}
\end{eqnarray}
%
is the ``high energy'' term and
\begin{eqnarray}
A_{\rm MIX} = & - & \left(\frac{\dmsol}{\dma}\right )^{\frac{1}{4}}\, F_1\, 
|U_{\t2}^*U_{\t3}|\, 
\left [ \, \eta(0.66\widetilde{m}_\tau) -
\eta(0.71\widetilde{m}_2)\, \right ] 
\nonumber\\
& \times & 
\left[\,\sin(\tilde{\f}_{12} + \tilde{\f}_{13} + \frac{\a_{32}}{2}) + 
\left(\frac{\dmsol}{\dma}\right)^{\frac{1}{2}}\,  
\sin(\tilde{\f}_{12} + \tilde{\f}_{13} - \frac{\a_{32}}{2})
\right]\,,\nonumber\\
\label{AMIX}
\end{eqnarray}
%
is the ``mixed'' term. Note that
for the best fit value of 
$s^2_{23} = 0.5$, we have $|U_{\t3}|^2 = c^2_{23}c^2_{13} 
\cong 0.5 \cong (1 - |U_{\t3}|^2)^{1/2}$, and 
therefore effectively $A_{\rm HE} \propto
(\eta(0.66\widetilde{m}_\tau) + \eta(0.71\widetilde{m}_2))$.
For $\tilde{\f}_{12}=k\pi/2$, $\tilde{\f}_{13}=k'\pi/2$, 
$k,k'=0,1,2,...$, one has $A_{\rm HE} = 0$ and 
we recover the expression for $Y_B$ from \cite{PPRio106}
when the only source of CP-violation are the Majorana 
phases in the PMNS matrix $U$. 
One can have successful leptogenesis 
in this case typically for 
$M_1 \gtap 4\times 10^{10}$ GeV  and
$|\sin \a_{32}/2| \gtap 0.1$ \cite{PPRio106}.
The phase $\a_{32}$ is also present in the 
expression for the neutrinoless double beta 
($\betabeta$-) decay effective Majorana mass
corresponding to the NH spectrum
(see, e.g. \cite{BPP1,STPFocusNu04}).
 
 Few more comments are in order.
It follows from eq. (\ref{YBMaj1})
that the $\tau$ and $(e + \mu)$ 
CP-violating asymmetries 
generated by the ``high energy'' term 
always add up, while the  
$\tau$ and $(e + \mu)$  asymmetries 
due to the ``mixed'' term tend to 
compensate each other. The contribution 
of the ``mixed'' term to $Y_B$ 
has the additional ``suppression''
factor $(\dmsol/\dma)^{1/4}\cong 0.42$
in comparison to that due to the 
``high energy'' term. 
For $\sin(\tilde{\f}_{12} + \tilde{\f}_{13} + \a_{32}/2) = 0$, 
the ``mixed'' term $|A_{\rm MIX}|$, 
as can be shown, 
is smaller at least by the factor 
$(\dmsol/\dma)^{1/2}c_{12}/\sqrt{2} \cong 0.11$ than 
the ``high energy'' term $|A_{\rm HE}|$. 
Finally, the sign of $A_{\rm HE}$ is determined 
by the sign of $\sin 2\tilde{\f}_{13}$, 
while the sign of $A_{\rm MIX}$
depends on the signs of 
$\sin(\tilde{\f}_{12} + \tilde{\f}_{13} + \a_{32}/2)$
and $(\eta(0.66\widetilde{m}_\tau) - \eta(0.71\widetilde{m}_2))$.

  The ``high energy'' term 
$A_{\rm HE}\propto F_{3}\sin 2\tilde{\f}_{13}$ 
will be suppressed  and will give a subdominant 
contribution in 
$|Y_B|$ if either the phase of $R^2_{13}$ 
is to a good approximation CP-conserving so that 
$\sin 2\tilde{\f}_{13} \cong 0$,
or $|R_{13}|/|R_{12}|$ is sufficiently small.
For $\sin 2\tilde{\f}_{13} = \sin(\f_{23} + \bar{\f}_{23}) = 0$ and
$|R_{13}|,|R_{12}|\neq 0$, however, we also have 
$\sin(\f_{23} - \bar{\f}_{23}) = 0$, implying that 
$R^2_{12}$ and $R^2_{13}$ are real, while
$R_{12}R_{13}$ is real or purely imaginary.
This case has been studied in detail in 
\cite{PPRio106,MPST07}. 
If, on the other hand, $|R_{13}|=0$, we will 
have $\eps_\t = \eps_2 = 0$, and, 
as a consequence, $Y_B=0$. Thus, in order 
to have successful leptogenesis 
in the case of interest, the ratio
$|R_{13}|/|R_{12}|$ should not be too small, 
i.e. should be larger than approximately 0.05. 

 For the wash-out mass parameter 
$\mtil_{\t}$ we have:
\begin{eqnarray}
\mtil_{\t} & = & \sqrt{\dmsol}\,|R_{12}|^2\,|U_{\t2}|^2\, 
+ \,\sqrt{\dma}\,|R_{13}|^2\,|U_{\t3}|^2\nonumber\\ 
& - & 2\,(\dmsol\,\dma)^{1/4}\,|R_{12}|\,|R_{13}|\,
|U_{\t2}^*U_{\t3}|\,
\cos \left (\,\tilde{\f}_{12} - \tilde{\f}_{13} + \frac{\a_{32}}{2} \right )\,.
\label{tmtNHMaj1}
\end{eqnarray}
%
Thus, for given $|R_{12}|$ and $|R_{13}|$,
$\mtil_{\t}$ 
satisfies the following inequalities: 
\begin{eqnarray}
\mtil_{\t} & \geq & \,\sqrt{\dma}\,|R_{13}|^2\,|U_{\t3}|^2\,
\left (1 - \left (\frac{\dmsol}{\dma}\right )^{1/4}\,
\frac{|R_{12}|}{|R_{13}|}\,\frac{|U_{\t2}|}{|U_{\t3}|}\right )^2\,, 
\label{mintmtMaj1}
\\ 
\mtil_{\t} & \leq & \,\sqrt{\dma}\,|R_{13}|^2\,|U_{\t3}|^2\,
\left (1 + \left (\frac{\dmsol}{\dma}\right )^{1/4}\,
\frac{|R_{12}|}{|R_{13}|}\,\frac{|U_{\t2}|}{|U_{\t3}|}\right )^2\,. 
\label{maxtmtMaj1}
\end{eqnarray}
%
It follows from eq. (\ref{tm21}) that 
the minimum (maximum) value of $\mtil_{\t}$ 
corresponds to the 
maximum (minimum) value of $\mtil_2$. 

 We have seen that for fixed $M_1$ and given values of the 
neutrino oscillations parameters, the asymmetry 
$Y_B$ and the relative contributions to $Y_B$ 
of the ``high energy'' and the ``mixed'' terms
depend on $|R_{12}|$, $|R_{13}|$ and the Majorana 
phase $\alpha_{32}$, or equivalently, 
on the three phases $\tilde{\f}_{12}$, $\tilde{\f}_{13}$
and $\alpha_{32}$. 
One of the constraints $\tilde{\f}_{12}$ and 
$\tilde{\f}_{13}$ should satisfy, as we have 
already indicated, is
 ${\rm sgn}(\sin 2\tilde{\f}_{12}) 
= - {\rm sgn}(\sin 2\tilde{\f}_{13})$ 
(see eq. (\ref{orth2})). It follows from 
eqs. (\ref{YBMaj1}) - (\ref{AMIX})  
and (\ref{tmtNHMaj1}) that 
$Y_B(\tilde{\f}_{12},\tilde{\f}_{13};
\alpha_{32}) = - Y_B(-\tilde{\f}_{12},-\tilde{\f}_{13};
4\pi - \alpha_{32})$. Therefore in what follows 
we shall  analyse only the case of 
$\sin 2\tilde{\f}_{12} < 0$, 
$\sin 2\tilde{\f}_{13} > 0$. 
The results corresponding to 
$\sin 2\tilde{\f}_{12} > 0$, 
$\sin 2\tilde{\f}_{13} < 0$ can be 
obtained from those we will derive 
using the indicated property of $Y_B$. 
In what concerns the values of 
$|R_{12}|$ and $|R_{13}|$, 
there are several possibilities leading to 
quite different physics results:\\ 
i) $|R_{13}| \leq |R_{12}|$
with $|R_{12}| \leq 1$;
ii) $|R_{12}| \leq |R_{13}|$
with $|R_{13}| \leq 1$;
iii) $|R_{12}| > 1$ 
or $|R_{13}| > 1$. 

  Consider first the case of 
$|R_{13}| \leq |R_{12}| \leq 1$. 
Under these conditions one always has 
$|R_{12}| \geq 1/\sqrt{2}$.
As we have already indicated earlier,
the asymmetry $|Y_B|$ will be strongly 
suppressed if $|R_{13}|/|R_{12}| \ll 0.05$, 
so we will limit our discussion 
to values of $|R_{13}|/|R_{12}| \gtap 0.05$.
The results we obtain depend on 
whether $|R_{13}| \ltap 0.5$ or 
$|R_{13}| \gtap 0.5$. 

  For $|R_{13}| \leq 0.5$, one should 
have $|R_{12}| > \sqrt{0.75} \cong 0.87$ in order 
for  $\sin2\tilde{\f}_{13} \neq 0$.
Let us set $|R_{12}| = 1$ for 
concreteness. In this case 
we get $\cos2\tilde{\f}_{12} = 1 - 0.5|R_{13}|^4 
\gtap 0.97$,
$|\sin2\tilde{\f}_{12}| = |R_{13}|^2\leq 0.25$, 
$\cos2\tilde{\f}_{13} = 0.5|R_{13}|^2\leq 0.125$, 
$|\sin2\tilde{\f}_{13}| \cong
1 - |R_{13}|^4/8 \gtap 1 - 7.8\times 10^{-3}$.
Thus, $ 0 < (- \tilde{\f}_{12}) \ltap 0.12$
and $\tilde{\f}_{13} \cong \pi/4$.
This implies that for 
$\alpha_{32}/2 \cong \pi/4$ we would have 
$\sin(\tilde{\f}_{12} + \tilde{\f}_{13} + \a_{32}/2) \cong 1$,
while if $\alpha_{32}/2 \cong 3\pi/4$
the ``mixed'' term will be strongly suppressed.
It follows from these simple observations that 
the predictions for $|Y_B|$ will exhibit a strong 
dependence on $\alpha_{32}$.
For $\alpha_{32}/2 \cong \pi/4$ we also have 
$\cos(\tilde{\f}_{12} - \tilde{\f}_{13} + \a_{32}/2)
\cong \cos\tilde{\f}_{12} \cong 1$, and for any given 
$|R_{13}|\leq 0.5$, $\mtil_{\t}$ will take 
to a good approximation its minimal value. 

  Let us analyse the behavior of 
$A_{\rm MIX}$ and $A_{\rm HE}$ for 
$\alpha_{32}/2 = \pi/4$ 
as $|R_{13}|$ decreases starting from the 
value $|R_{13}|= 0.5$ (Fig. 1). In this 
analysis we will set $\sin\theta_{13} = 0$ 
for simplicity and will use the best fit values  
of the other neutrino oscillation parameters 
given in eq. (\ref{deltaatmvalues});
we will comment on the possible effects 
of $\sin\theta_{13}$ having a value close to 
the existing upper limit later.
It is useful also to recall that 
the efficiency factor $\eta(X)$ 
i) has an absolute maximum at 
$X \cong 10^{-3}$ eV, and at the 
maximum $\eta \cong 7\times 10^{-2}$, 
and that ii) both for 
increasing $X > 10^{-3}$ eV and 
decreasing $X < 10^{-3}$ eV, 
$\eta(X)$ is a monotonically decreasing 
function of $X$.

   At $|R_{13}|= 0.5$ we have: 
$\mtil_{\t} \cong 5.7\times 10^{-4}$ 
(weak wash-out), 
$\mtil_2 \cong 2.1\times 10^{-2} \gg \mtil_{\t}$ 
(strong wash-out), 
$\eta(0.66\widetilde{m}_\tau) \cong 4.2\times 10^{-2}$,
and $\eta(0.71\widetilde{m}_2) \cong 6.8\times 10^{-3} 
< \eta(0.66\widetilde{m}_\tau)$. 
The ``mixed'' term and the  ``high energy'' term 
have opposite signs and 
$A_{\rm MIX} \cong - 7.3\times 10^{-3}$ 
and $A_{\rm HE}\cong 1.40\times 10^{-2}$.
Therefore the contribution of the ``mixed'' term 
in $Y_B$ has the effect of partially compensating 
the contribution of the ``high energy'' term, 
so that the sum $(A_{\rm MIX} + A_{\rm HE})$
is approximately by a factor of 2 smaller than 
$A_{\rm HE}$. As $|R_{13}|$ decreases starting from 0.5, 
$\mtil_{\t}$, $\mtil_2$ and $\eta(0.66\widetilde{m}_\tau)$ 
also decrease starting from the values given above. 
However, $\eta(0.71\widetilde{m}_2)$ increases 
and at $|R_{13}| \cong 0.41$~ 
\footnote{This value is obtained as a solution 
of the equation $0.66\mtil_{\t}/(8.25\times 10^{-3}~eV) 
= (0.71 \mtil_2/(2\times 10^{-4}~eV))^{-1.16}$.} 
one has $\eta(0.66\widetilde{m}_\tau) \cong 
\eta(0.71\widetilde{m}_2)$. As a consequence, 
at $|R_{13}| \cong 0.41$, 
$|A_{\rm MIX}|$ goes through a deep minimum
(at which it can even be 0) and is strongly suppressed. 
The ``high energy'' term
$|A_{\rm HE}|$ just decreases somewhat as 
$|R_{13}|$ changes from 0.50 to 0.41. 
At $|R_{13}| \cong 0.41$, 
the mixed term $A_{\rm MIX}$ 
changes sign: in the interval 
$|R_{13}| \cong (0.3 - 0.4)$ 
one has $\eta(0.66\widetilde{m}_\tau) < 
\eta(0.71\widetilde{m}_2)$ and, 
consequently $A_{\rm MIX} >0$.
Thus,  $A_{\rm HE}$ and $A_{\rm MIX}$ 
have the same sign and
add up constructively in 
$Y_B$. As $|R_{13}|$ decreases below
0.41, $\mtil_{\t}$, $\mtil_2$ 
and $\eta(0.66\widetilde{m}_\tau)$
continue to decrease, while 
$\eta(0.71\widetilde{m}_2)$
continues to increase; 
$A_{\rm MIX}$ also increases 
rapidly, while $A_{\rm HE}$
decreases but rather slowly (Fig. 1).
At $|R_{13}| \cong 
(\dmsol/\dma)^{1/4}c_{12} \cong 0.35$ 
we get $\mtil_{\t} \cong 0$ and 
$A_{\rm MIX}$ has a local maximum. 
At this point we have   
$A_{\rm MIX} \cong A_{\rm HE} \cong 2\times 10^{-3}$.
As $|R_{13}|$ decreases further, 
$\mtil_{\t}$ and $\eta(0.66\widetilde{m}_\tau)$ 
increase, $\widetilde{m}_2$ decreases, but
$\eta(0.71\widetilde{m}_2)$ increases.
As a consequence,  $A_{\rm HE}$ also increases, 
while $A_{\rm MIX}$ diminishes. 
At $|R_{13}| \cong 0.27$ we have   
$\eta(0.66\widetilde{m}_\tau)
\cong \eta(0.71\widetilde{m}_2)$ 
and $A_{\rm MIX}$ exhibits a second deep minimum,
$A_{\rm MIX} \cong 0$. At values of
$|R_{13}| < 0.27$ the inequality 
$\eta(0.66\widetilde{m}_\tau) >
\eta(0.71\widetilde{m}_2)$ holds 
and $A_{\rm MIX}$ is negative, 
$A_{\rm MIX} < 0$. Therefore 
$A_{\rm HE}$ and $A_{\rm MIX}$
have opposite signs and their 
contributions to $Y_B$ tend to compensate 
each other. For decreasing  $|R_{13}| < 0.27$,
$\eta(0.66\widetilde{m}_\tau)$ and 
$F_1 (\eta(0.66\widetilde{m}_\tau) - 
\eta(0.71\widetilde{m}_2))$ 
grow faster than $\eta(0.71\widetilde{m}_2)$
and $F_3 (\eta(0.66\widetilde{m}_\tau) + 
\eta(0.71\widetilde{m}_2))$, respectively.
At  $|R_{13}| \cong 0.18$, 
$A_{\rm HE}$ has a local maximum. 
However, one also has 
$|A_{\rm MIX}| \cong A_{\rm HE}$. 
As a consequence, $A_{\rm MIX} + A_{\rm HE} \cong 0$, 
i.e. the ``high energy'' and the ``mixed'' terms 
cancel each other and $|Y_B|$ is strongly 
suppressed. This important feature of 
$|Y_B|$ persists for values of 
$\alpha_{32}/2$ up to  $\sim \pi/2$.
The precise position of the 
considered deep minimum of $|Y_B|$ 
depends on the value of 
$\sin^2\theta_{23} \equiv s^2_{23}$ and, 
to less extent, on whether $\delta = 0$ or $\pi$ if 
$\sin\theta_{13} \equiv s_{13}$ has a 
value close to the existing upper limit. 
As an illustration, in Fig. 2 we show 
$|Y^0_B A_{\rm HE}|$, $|Y^0_B A_{\rm MIX}|$ and 
$|Y_B|$ as functions of $|R_{13}|$ for $s^2_{23} = 0.64$,
$s_{13} = 0.2$ and $\delta = 0$. 
As is seen in the figure, for $s^2_{23} = 0.64$ 
we have $A_{\rm MIX} + A_{\rm HE} \cong 0$, and 
correspondingly $Y_B\cong 0$, at
$|R_{13}| \cong 0.30$. Note that 
both $|Y^0_B A_{\rm HE}|$ and $|Y^0_B A_{\rm MIX}|$
have relatively large values at $|R_{13}| \cong 0.30$  
and thus each of the two terms separately could 
account for the observed value of $Y_B$ (Fig. 2). 
Nevertheless, the generated baryon asymmetry
is strongly suppressed, 
$|Y_B| = |Y^0_B (A_{\rm HE}+ A_{\rm MIX})| \ll 8.6\times 10^{-11}$ 
and it is impossible to reproduce the measured value of 
$Y_B$ for $M_1 \ltap 10^{12}$ GeV.

  For $|R_{13}| < 0.17$, the ``mixed'' term is 
larger (in absolute value) 
than the ``high energy'' term, $|A_{\rm MIX}| > A_{\rm HE}$; 
at $|R_{13}| = 0.10$, for instance, we have 
$|A_{\rm MIX}| \cong 2A_{\rm HE}$. 
Since the two terms have opposite signs, 
${\rm sgn}(A_{\rm MIX}) = - {\rm sgn}(A_{\rm HE})$, 
the contributions of the ``high energy'' term 
in $Y_B$ compensates partially the contribution 
of the ``mixed'' term. 

  We have studied also the dependence of 
the baryon asymmetry $|Y_B|$
on the Majorana phase $\alpha_{32}$
for given fixed values of $|R_{13}| \leq |R_{12}| = 1$
from the interval $0.1 \ltap |R_{13}| \ltap 0.5$.
This was done for three values of 
$s^2_{23} = 0.36;~0.50;~0.64$ and 
two values of $s_{13} = 0;~0.2$.
In the case of $s_{13} = 0.2$, 
the two CP-conserving values of the Dirac 
phase $\delta$ were considered: $\delta =0;~\pi$. 
These results are illustrated in Figs. 3 - 5.
As these figures indicate, 
the behavior of 
$|Y_B|$ as a function of $\alpha_{32}$
exhibits particularly interesting features 
when $\alpha_{32}$ changes in the interval
$0 < \alpha_{32} \ltap \pi$.
We note only that  for $s_{13} = 0.2$ and given 
$s^2_{23}$, we can get very 
different dependence of 
$|Y_B|$ on $\alpha_{32}$
for the two values of $\delta =0;~\pi$ 
(Fig. 3), and that
the dependence under discussion
for, e.g. $s^2_{23} = 0.50$
can differ drastically from those 
for $s^2_{23} = 0.36$  
and for $s^2_{23} = 0.64$ (Figs. 4 - 5). 

  We can analyse in a similar manner 
the behavior of $A_{\rm MIX}$, 
$A_{\rm HE}$ and $|Y_B|$ as functions $|R_{13}|$ 
in the interval $0.5 <  |R_{13}| \leq 1.0$. 
As in the preceding discussion we assume that
$|R_{13}| \leq |R_{12}|\leq 1.0$ and fix again 
$\alpha_{32}/2 = \pi/4$.  
As can be easily verified, 
when $|R_{13}|$ increases from 0.5 to 1.0 under
the indicated conditions, 
i) $F_1\sin(\tilde{\f}_{12} + \tilde{\f}_{13} + \a_{32}/2)$ 
changes from 1.14 to 0.60,
ii) $F_3 \sin 2\tilde{\f}_{13}$ increases from 
0.59 to 0.74,
iii) $\widetilde{m}_\tau$ increases monotonically 
by a factor of $\sim 20$ 
from $5.7\times 10^{-4}$ eV to $1.1\times 10^{-2}$ eV, 
and iv) $\widetilde{m}_2$ increases only by a factor of 
$\sim 2.3$ from $2.1\times 10^{-2}$ eV to $4.8\times 10^{-2}$ eV.
Correspondingly, the efficiency factor
$\eta(0.66\widetilde{m}_\tau)$ first increases
starting from the value $4.2\times 10^{-2}$, 
reaches a maximum 
$\eta(0.66\widetilde{m}_\tau) \cong 6.8\times 10^{-2}$ 
at $|R_{13}| \cong 0.6$ when
$0.66\widetilde{m}_\tau \cong 1.1\times 10^{-3}$ eV,
and then decreases 
monotonically to $1.52\times 10^{-2}$.
In contrast, when $|R_{13}|$ changes from 0.5 to 1.0,
the efficiency factor $\eta(0.71\widetilde{m}_2)$
only decreases monotonically by a factor of $\sim 2.6$ 
approximately from $6.7\times 10^{-3}$ to $2.6\times 10^{-3}$.  
Thus, the asymmetry in the $(e +\mu)$ lepton charge 
is generated in the regime of strong wash-out, 
while the wash-out effects in the 
production of the asymmetry in $\tau$ 
lepton charge change from weak to 
strong passing through a minimum. 
Clearly, the change of 
$A_{\rm MIX}$ and $A_{\rm HE}$ 
with $|R_{13}|$ is determined essentially 
by the behavior of $\eta(0.66\widetilde{m}_\tau)$.
We  also have $\eta(0.66\widetilde{m}_\tau) > 
\eta(0.71\widetilde{m}_2)$ in the case  
under discussion, implying that 
${\rm sgn}(A_{\rm MIX}) = - {\rm sgn}(A_{\rm HE})$.
For the considered range of $|R_{13}|$ one 
typically has $|A_{\rm MIX}| \cong (0.5 - 0.6)A_{\rm HE}$, 
so there is a partial cancellation between 
the two terms $A_{\rm MIX}$ and $A_{\rm HE}$ 
in $Y_B$ (Fig. 1).

   It should be clear that 
$A_{\rm MIX}$, $A_{\rm HE}$ and $|Y_B|$
will exhibit a different dependence 
on $|R_{13}|$ varying in the range
$0.05\ltap |R_{13}|\leq |R_{12}| \leq 1$ 
if $\alpha_{32}/2$ differs significantly 
from  $\pi/4$. 
If $\alpha_{32}/2 \cong 3\pi/4$,
for instance, we get typically 
$|A_{\rm MIX}| \ll |A_{\rm HE}|$.
For $|R_{13}| \ltap 0.5$ this essentially 
is due to the fact that 
$\sin(\tilde{\f}_{12} + \tilde{\f}_{13} + \a_{32}/2) \ll 1$, 
while for $0.5 < |R_{13}| \leq 1$, 
$|R_{12}| \cong 1$, it is a consequence of 
the fact that $\eta(0.66\widetilde{m}_\tau)$ and 
$\eta(0.71\widetilde{m}_2)$ have rather close values:
when $|R_{13}|$ changes from 0.5 to 1.0,
$\eta(0.66\widetilde{m}_\tau) - \eta(0.71\widetilde{m}_2)$ 
decreases approximately from 
$7.6\times 10^{-3}$ to $2.7\times 10^{-3}$; 
at the same time the sum 
$\eta(0.66\widetilde{m}_\tau)+ \eta(0.71\widetilde{m}_2)$
changes from $3\times 10^{-2}$ to $10^{-2}$, 
remaining by a factor $\sim 4$
bigger than $(\eta(0.66\widetilde{m}_\tau) - \eta(0.71\widetilde{m}_2))$.

  One can perform a similar analysis in the case of
$|R_{12}| > 1$ or $|R_{13}| > 1$. The results we have 
obtained for $|R_{12}| > 1$ are illustrated in Fig. 6, 
which shows the dependence of $|Y^0_B A_{\rm HE}|$,
$|Y^0_B A_{\rm MIX}|$ and of $|Y_B|$  
on $|R_{13}|$ for $|R_{12}|=1.2$,
$\alpha_{32}/2 = \pi/4$ and
$s_{23}^2=0.5$, $s_{13}=0$, $M_1=10^{11}\GeV$.
The figure exhibits some typical features, 
namely, the relevance of the ``mixed'' term
in the region close to the 
minimal allowed value of $|R_{13}|$, 
i.e. for  $|R_{13}| \ltap 1$.
If $|R_{12}| > 1$ (say $|R_{12}| = 1.2$
as in Fig. 6), $|R_{13}|^2$ can take values 
in the interval $(|R_{12}|^2 - 1) \leq 
|R_{13}|^2 \leq (|R_{12}|^2 + 1)$.
When $|R_{13}|^2$ changes from its minimal value 
to its maximal value, the phase 
$2\tilde{\f}_{13}$, as can be easily shown, 
decreases from $\pi$ to 0, whereas 
$2\tilde{\f}_{12}$ changes from 
0 to $(-\pi)$, so that 
one always has $\sin 2\tilde{\f}_{12} \leq 0$.
Obviously, at $|R_{13}|^2 = (|R_{12}|^2 - 1)$ 
we have $A_{\rm HE} = 0$ since 
$\sin 2\tilde{\f}_{13} = 0$, while 
for $\alpha_{32}/2 \neq \pi k$, $k=0,1,2,...$, 
one finds, in general, $A_{\rm MIX} \neq 0$.
For the value of $\alpha_{32}/2 = \pi/4$ 
(Fig. 6), for instance, we get
$A_{\rm MIX} \cong - 3.9\times 10^{-3}$.
The salient features of the behavior 
of $A_{\rm HE}$ and $A_{\rm MIX}$ 
as functions of $|R_{13}|$ shown in Fig. 6,
can be understood qualitatively from the 
behavior essentially of 
$F_3\sin (2\tilde{\f}_{13})\eta(0.66\widetilde{m}_\tau)$ and of
$F_1\sin (\tilde{\f}_{12}+ \tilde{\f}_{13}+\alpha_{32}/2) 
\eta(0.66\widetilde{m}_\tau)$: both quantities 
first grow relatively fast monotonically 
as $|R_{13}|$ increases starting from 
its minimal value, but the former 
grows faster that the latter. 
Since i) $A_{\rm HE}$ increases 
faster than $|A_{\rm MIX}|$ starting from 0, 
ii) except at the extreme values of $|R_{13}|$ where 
$A_{\rm HE} = 0$,
we have $A_{\rm HE} > 0$, 
iii) $|A_{\rm MIX}|$ increases 
starting from a finite value
but $A_{\rm MIX} < 0$, 
there is always a value of 
$|R_{13}|$ relatively close to its minimal value
at which $A_{\rm HE} = |A_{\rm MIX}|$.  
Obviously, at this point
the baryon asymmetry is strongly 
suppressed: $Y_B = Y^0_B(A_{\rm HE} + A_{\rm MIX}) =0$ (Fig. 6).
The behavior of $A_{\rm HE}$ and $|A_{\rm MIX}|$ as 
$|R_{13}|$ increases beyond the point at 
which $Y_B\cong 0$, is basically determined by 
$\eta(0.66\widetilde{m}_\tau)$, which goes 
through a maximum and after that decreases monotonically.
Let us note also that at certain value of
$|R_{13}| > 1$,  
$\sin (\tilde{\f}_{12}+ \tilde{\f}_{13}+\alpha_{32}/2)$ 
can go through zero and changes sign.
As a consequence, $A_{\rm MIX}$ also can change sign.

   As the results described above show, in the case of 
NH light neutrino mass spectrum and CP violation due the 
``low energy'' Majorana phases in $U_{\rm PMNS}$
and ``high energy'' $R$-phases, the 
predicted baryon asymmetry can exhibit
strong dependence on the Majorana phase $\alpha_{32}$
if the latter has a value 
in the interval $0 < \alpha_{32} < \pi$
($\sin 2\tilde{\f}_{12} < 0$, $\sin 2\tilde{\f}_{13}> 0$), 
or  $3\pi < \alpha_{32} < 4\pi$
($\sin 2\tilde{\f}_{12} > 0$, $\sin 2\tilde{\f}_{13}< 0$).
In the most extreme cases we can have either
$Y_B = 0$ or $Y_B$ compatible with the observations 
in a certain point of the relevant parameter space,
depending on the value of $\alpha_{32}$.

%
\subsection{CP Violation Due to Dirac Phase in $U_{\rm PMNS}$
and  $R-$Phases}
%
%
Consider next the possibility that
the CP-violation in ``flavoured'' leptogenesis
is due to the Dirac phase $\d$ in the PMNS matrix $U$
and to the ``high energy'' phases $\tilde{\f}_{12}$ and
$\tilde{\f}_{13}$ of the matrix $R$.
The Majorana phase $\a_{32}$ will be assumed to
take a CP-conserving value:
$\a_{32} =\pi k$, $k=0,1,2,...$.
The expression for the baryon asymmetry $Y_B$
also in this case can be cast in the form (\ref{YBMaj1}).
The ``high energy'' term $A_{\rm HE}$ is the same
as in the Majorana and R-matrix CP-violation
case and is given by eq. (\ref{AHE1}).
The ``mixed'' term has the following form for arbitrary
$\a_{32}$:
\begin{eqnarray}
A^{\rm D}_{\rm MIX} & = & 
- \,\left(\frac{\dmsol}{\dma}\right)^{1/4}\,F_1\,c_{23}\,c_{13}
\,[\eta(0.66\mtil_\t)-\eta(0.71\mtil_2)]\nonumber\\\\
& \times &
\left\{c_{12}s_{23}\left(
\sin\left(\tilde{\f}_{12}+\tilde{\f}_{13} +
\frac{\a_{32}}{2}\right)+
\sqrt{\frac{\dmsol}{\dma}}\sin\left(\tilde{\f}_{12}+
\tilde{\f}_{13}-\frac{\a_{32}}{2}\right)\right)
+\Phi_{\rm MIX}^{\rm D}\right\}\,,\nonumber
\label{NHDCPAmix1}
\end{eqnarray}
%
where
\begin{equation}
\Phi_{\rm MIX}^{\rm D}\,=\,
s_{12}c_{23}s_{13}
\left[\sin\left(\tilde{\f}_{12}+\tilde{\f}_{13}+
\frac{\a_{32}}{2}-\d\right)
+\sqrt{\frac{\dmsol}{\dma}}\sin\left(\tilde{\f}_{12}+
\tilde{\f}_{13}
- \frac{\a_{32}}{2}+\d \right)\,\right]\,.
\label{NHDterm1}
\end{equation}
%

The wash out mass parameter $\mtil_\t$ is given by
\begin{eqnarray}
\mtil_{\t} & = & \sqrt{\dmsol}\,|R_{12}|^2\,|U_{\t2}|^2\,
+ \,\sqrt{\dma}\,|R_{13}|^2\,|U_{\t3}|^2
- 2\,(\dmsol\,\dma)^{1/4}\,|R_{12}|\,|R_{13}|\,
c_{23}c_{13}\,\nonumber\\
& \times &
\left[\,c_{12}s_{23}\,
\cos\left(\,\tilde{\f}_{12}- \tilde{\f}_{13}+
\frac{\a_{32}}{2}\,\right) + s_{12}c_{23}
s_{13}\cos\left(\,\tilde{\f}_{12}-\tilde{\f}_{13}+
\frac{\a_{32}}{2}-\d\,\right)\,\right]\,.
\end{eqnarray}
%

 For e.g.  $\a_{32}= 2\pi k$, $k=0,1,2,...$,
and $\tilde{\f}_{12}, \tilde{\f}_{13}=0,\pm \pi$,
$R_{12}$ and  $R_{13}$ are real,
$A_{\rm HE} = 0$, while
in the ``mixed'' term only
the part proportional to
$\Phi_{\rm MIX}^{\rm D}$ is non-zero,
$A^{\rm D}_{\rm MIX} \propto \Phi_{\rm MIX}^{\rm D} \neq 0$.
The CP-violation in leptogenesis in this case
is  entirely due to the Dirac phase
$\delta$ in the PMNS matrix and we
recover the results obtained in \cite{PPRio106}.
In particular, one can have successful leptogenesis
for $M_1 \ltap 5\times 10^{11}$ GeV
provided $|s_{13}\sin \delta|\gtap 0.1$
~\footnote{Values of $s_{13} \gtap 0.1$ are within 
the range to be probed by 
``near'' future experiments 
with reactor $\bar{\nu}_e$ \cite{DCHOOZ,DayaB}.
Future long baseline experiments
will aim at measuring values of $\sin^2\theta_{13}$ 
as small as $10^{-4}$--$10^{-3}$ and at constraining 
(or determining) $\delta$~(see, e.g. \cite{Future}).}.
For $\a_{32} = 0$ and $R_{12}R_{13} > 0$
($R_{12}R_{13} < 0$),
the baryon asymmetry $|Y_B|$
has a maximum at $R^2_{12} \cong 0.75$, $R^2_{13} \cong 0.25$
($R^2_{12} \cong 0.85$, $R_{13}^2 \cong 0.15$).

 We are interested in the regions
of parameter space where
there is a noticeable interplay
between the CP violation due to the
``high energy'' phases $\tilde{\f}_{12}$ and
$\tilde{\f}_{13}$ and
the CP violation due to the Dirac
phase $\delta$.
Since the CP-violation effects due to the Dirac
phase are always suppressed by the relatively small
experimentally allowed value of $s_{13}$,
the regions of interest would correspond to
$\tilde{\f}_{13}\sim 0,\,\pm\pi/2$,
where $A_{\rm HE}$ is also suppressed.
The case of $\tilde{\f}_{13}\sim 0,\,\pm\pi/2$,
corresponds to $|R_{13}|$ taking values
close to the boundaries:
$|R_{13}|^2\sim |\,|R_{12}|^2 \mp 1\,|$.

  Note that the ``mixed'' term $A^{\rm D}_{\rm MIX}$
contains a piece which does not depend on the
Dirac phase $\delta$. This $\delta$-independent piece
is multiplied by $c_{12}s_{23}$
which is approximately at least by a factor 7 larger
than the corresponding mixing angle factor
$s_{12}c_{23}s_{13}$
in the $\delta$-dependent term
$\Phi_{\rm MIX}^{\rm D}$. In the region
$|R_{13}|^2\sim |\,|R_{12}|^2 \mp 1\,|$,
we also have $\sin (\tilde{\f}_{12} +\tilde{\f}_{13}+\a_{32}/2)\cong 0$
for $\a_{32}= \pi k$, and the $\delta$-independent term
in $A^{\rm D}_{\rm MIX}$ will also be suppressed.

  We have performed a detailed numerical analysis
of this region of parameter space
for CP-violating values of the Dirac phase
$\delta$ and a CP-conserving
Majorana phase  $\alpha_{32}$.
The results of this analysis show that
successful leptogenesis can still be realized for
$|R_{13}|^2\gtap |\,|R_{12}|^2-1\,|$ and 
$|R_{12}|\approx\mathcal{O}(1)$. Moreover, 
in the cases we have considered, the effects of the CP-violating
Dirac phase are relevant in order to reproduce the observed
value of the baryon asymmetry.


    In Fig. \ref{NH_YB_delta} we show
$|Y_B|$ as a function of $\delta$ for $|R_{12}|\cong 1$,
$s_{13}=0.2$, i) $\alpha_{32}=0$ (left panel) and
ii) $\alpha_{32}=\pi$ (right panel). We choose $|R_{13}|$ close
to its lower bound. In both the shown cases, there is a significant 
interference between the ``high energy'' 
and the ``mixed'' terms that can 
suppress or enhance the baryon
asymmetry. The latter is controlled 
by the Dirac phase $\d$.

   Thus, it follows from our analysis
that if the Majorana phase $\alpha_{32}$
possesses a CP conserving value,
there still will be regions in the parameter
space where the effects of the
CP-violating Dirac phase in the PMNS matrix
can be significant in ``flavoured'' leptogenesis,
even if CP-violation is due also to the ``high energy''
$R$-matrix phases.

%
\section{Inverted Hierarchical Light Neutrino Mass Spectrum}
%
%
 We get very different results in the case 
of IH light neutrino mass spectrum.  Now we have
$m_3\,\ll\,m_{1,2}\,\cong\,\sqrt{|\dma|}\,\cong\,0.05\,{\rm eV}$.
As in the analysis of NH mass spectrum, in what 
follows we neglect the effects of the lightest neutrino mass, $m_3$.
We also set $|R_{13}|=0$ ($N_3$ decoupling \cite{PRST05}) 
for simplicity. The orthogonality condition for 
the elements of the $R$ matrix of interest reads:
$R^2_{11} + R^2_{12} = 1$. It leads to 
constraints similar to those given in eqs. (\ref{orth1}) 
and (\ref{orth2}). They permit to express  
the high energy CP-violating phases relevant for the discussion 
in this Section, $\tilde{\f}_{11}$ and $\tilde{\f}_{12}$, 
in terms of the absolute values of $|R_{11}|$ and $|R_{12}|$:
\begin{eqnarray}
\cos 2\tilde{\f}_{11} & = &
\frac{ 1 + |R_{11}|^4 - |R_{12}|^4}{2|R_{11}|^2}\,,~~
\sin 2\tilde{\f}_{11}\,=\, \pm \sqrt{1 - \cos^2 2\tilde{\f}_{11}}\,,
\label{tef11} \\
\cos 2\tilde{\f}_{12} & = &
\frac{ 1 - |R_{11}|^4 + |R_{12}|^4}{2|R_{12}|^2}\,,~~
\sin 2\tilde{\f}_{12}\,=\, \mp \sqrt{1 - \cos^2 2\tilde{\f}_{12}}\,.
\label{tef12} 
\end{eqnarray}
%
\newline
The coefficients $|R_{11}|$ and $|R_{12}|$ are 
constrained by conditions similar to those given 
in eqs. (\ref{tf12R}) and (\ref{tf13R}); 
they can formally be obtained from the latter 
by replacing $|R_{13}|$ with $|R_{11}|$.
We shall have also $|R_{11}|^2\sin2\tilde{\f}_{11} 
+ |R_{12}|^2 \sin2\tilde{\f}_{12}=0$.
Using this condition one can write
the CP violating asymmetry $\eps_\t$ in the form:
\begin{eqnarray}\label{CP-asym-epstau_IH}
\eps_\t & \cong & 
-\frac{3\,M_1}{16\,\pi\, v^2}\,\frac{\sqrt{|\dma|}}{|R_{11}|^2\,
+\,|R_{12}|^2}\,\left\{|R_{11}|^2\,\sin(2\tilde{\f}_{11})\,
(|U_{\t1}|^2 - |U_{\t2}|^2)\right.\nonumber\\\\
& + &
\left.|R_{11}|\,|R_{12}|\left[\half\,\frac{\dmsol}{\dma}\,
\cos(\tilde{\f}_{11}+\tilde{\f}_{12})\,{\rm Im}(U_{\t1}^*U_{\t2})\,
+ \,2\,\sin(\tilde{\f}_{11}+\tilde{\f}_{12})
\,{\rm Re}(U_{\t1}^*U_{\t2})\right]\right\}\nonumber
\end{eqnarray}
%
For $\tilde{\f}_{11}=k\pi/2$,
$\tilde{\f}_{12}=k'\pi/2$, $k,k'=0,1,2,...$,
$R_{11}$ and $R_{12}$ are either real or purely 
imaginary and the expression for $\eps_\t$ 
reduces to the one derived in \cite{PPRio106}.
As was noticed in  \cite{PPRio106}, 
the asymmetry will be strongly suppressed 
(rendering successful leptogenesis impossible)
if $\sin(\tilde{\f}_{11}+\tilde{\f}_{12})=0$, 
i.e. if $R_{11}R_{12}$ is real;
if, however, $R_{11}R_{12}$ is purely imaginary, 
then $|\sin(\tilde{\f}_{11}+\tilde{\f}_{12})|=1$
and one can have successful leptogenesis 
with CP-violation due exclusively 
to the Majorana or Dirac phases in 
the PMNS matrix.

   It is easy to convince oneself 
using the expression for 
$\eps_\t$ given above and eq. (\ref{eps2}) 
that in spite of the presence of 
``high energy'' CP violation,
the following relation holds 
in the case under discussion: $\eps_2=-\eps_\t$. 
Thus, the baryon asymmetry $Y_B$ can be written 
as a function of $\eps_\t$ only, like in the 
case of the matrix $R$ satisfying the CP-invariance 
constraints, i.e. having real and/or 
purely imaginary elements \cite{PPRio106}:
\begin{equation}
Y_B\,=\,-\frac{12}{37}\frac{\eps_\t}{g_*}
\left(\, \y\left(\frac{390}{589}\mtil_{\t}\right)\, 
- \,\y\left(\frac{417}{589}\mtil_2\right)\right)\label{YB_2}
\end{equation}
%
For the wash-out mass parameters we obtain: 
\begin{eqnarray}
\mtil_{\t} & \cong & \sqrt{|\dma|}\,\left[\,|R_{11}|^2\,|U_{\t1}|^2\,
+\,|R_{12}|^2\,|U_{\t2}|^2
\right.\nonumber\\\\
 	   & + & \left. 2\,|R_{11}|\,|R_{12}|\,{\rm Re}
\left(e^{i\,(\tilde{\f}_{11} - \tilde{\f}_{12})}
\,U_{\t1}^*U_{\t2}\right)\,\right]
\nonumber\\\nonumber\\
\mtil_2 & = & \sqrt{|\dma|}\,(|R_{11}|^2\,+\,|R_{12}|^2)\,-\,\mtil_{\t}	   
\end{eqnarray}
%

%
\subsection{CP Violation Due to Majorana Phase in $U_{\rm PMNS}$ 
and  $R-$Phases}
%
%
We assume that the Dirac phase $\delta = k\pi$, $k=0,1,2,...$
The baryon asymmetry takes the form:
\begin{equation}
Y_B\,=\,Y^0_B\,(\,A^{\rm IH}_{\rm HE}\,+\,A^{\rm IH}_{\rm MIX}\,)
= \,Y^0_B\,(\,C_{\rm HE}\,+\,C_{\rm MIX}\,)
\left(\, \y\left(0.66\mtil_{\t}\right)\, - 
\,\y\left(0.71\mtil_2\right)\right)\,,
\label{YB_3}
\end{equation}
%
where $Y^0_B$ is given in eq. (\ref{Y0B}), 
$A^{\rm IH}_{\rm HE} \propto C_{\rm HE}$ is the 
``high energy'' term,
$A^{\rm IH}_{\rm MIX}\propto C_{\rm MIX}$ is the 
``mixed'' term,
\begin{eqnarray}
C_{\rm HE}\, = \, G_{11}\,  
\sin 2\tilde{\f}_{11} 
\left [\,|U_{\t1}|^2-|U_{\t2}|^2\,\right ]\,,
\label{CHE1}
\end{eqnarray}
%
\begin{eqnarray}
C_{\rm MIX} = -\,G_{12} 
|U_{\t1}^*U_{\t2}|\left[\frac{1}{2}\,
\frac{\dmsol}{|\dma|}\cos(\tilde{\f}_{11} + 
\tilde{\f}_{12})\sin\,\frac{\a_{21}}{2}
+2\sin(\tilde{\f}_{11}+\tilde{\f}_{12})\cos\,\frac{\a_{21}}{2}\,\right]
\label{CMIX1}
\end{eqnarray}
%
and $G_{11}\equiv|R_{11}|^2/(|R_{11}|^2+|R_{12}|^2)$, 
$G_{12}\equiv|R_{11}R_{12}|/(|R_{11}|^2+|R_{12}|^2)$. 
In deriving these expressions 
we have used the fact that for $\delta = \pi k$ 
and the values of the neutrino mixing angles 
allowed by the data we have
$(exp(-i\alpha_{21}/2)U_{\t1}^*U_{\t2}) = 
-[s_{12}c_{12}(s_{23}^2-c_{23}^2 s_{13}^2)
\pm s_{23}c_{23}c_{13}s_{13}(s_{12}^2-c_{12}^2)] = 
-|U_{\t1}^*U_{\t2}|$. 
Note that the first term in the square brackets 
in eq. (\ref{CMIX1}) is suppressed by the factor
$0.5 \dmsol/|\dma|\cong 0.016$, which renders it 
practically negligible in what concerns the 
discussion that follows. Note also that in contrast 
to the case of NH spectrum, both the ``high energy'' and the 
``mixed'' terms in $Y_B$ are proportional 
to the difference of the two relevant efficiency 
factors $(\y(0.66\mtil_{\t}) - \y(0.71\mtil_2))$.
The wash-out mass parameter $\mtil_\t$ now reads:
\begin{eqnarray}
\mtil_\t &=& \sqrt{|\dma|}\left[|R_{11}|^2|U_{\t1}|^2+ 
|R_{12}|^2|U_{\t2}|^2\right.\nonumber\\
&-&\left.2|R_{11}R_{12}||U_{\t1}^* U_{\t2}|
\cos\left(\tilde{\f}_{11}-\tilde{\f}_{12} + 
\frac{\a_{21}}{2}\right)\right]. 
\label{mtIHMaj1}
\end{eqnarray}
%
We see that both the ``mixed'' term $A^{\rm IH}_{\rm MIX}$
and the wash-out mass parameter $\mtil_\t$
(and thus the efficiency factor in $Y_B$)
depend on the Majorana phase $\a_{21}$.
The latter determines the range of 
possible values of the effective Majorana mass 
in neutrinoless double beta decay
in the case IH light neutrino mass spectrum 
\cite{STPFocusNu04}.

 Consider next the contributions of 
the ``mixed'' and the ``high energy'' terms to 
$Y_B$. We begin by noting that 
\begin{eqnarray}
|U_{\t1}|^2-|U_{\t2}|^2 &\cong& (s_{12}^2 - c_{12}^2)s_{23}^2-
4\,s_{12} c_{12} s_{23} c_{23} s_{13}\cos\d\nonumber\\
&\cong& -0.20 - 0.92\,s_{13}\,\cos\d\,,
\end{eqnarray}
%
where we have used $s_{12}^2=0.30$ and
$s_{23}^2 = 0.5$. Similarly, we have 
\begin{eqnarray}
|U_{\t1}^* U_{\t2}|&\cong& |s_{12} c_{12} s_{23}^2 -
s_{13} s_{23} c_{23}( c_{12}^2 - s_{12}^2 )\cos\d|\nonumber\\
&\cong& 0.50\,|0.46 - 0.40\,s_{13}\,\cos\d|\,.
\end{eqnarray}
%
Obviously, for $s_{13} = 0$ one obtains
$(|U_{\t2}|^2 - |U_{\t1}|^2) \cong |U_{\t1}^*U_{\t2}|$.
However, if $s_{13} = 0.2$ and $\delta = \pi$, 
we get $(|U_{\t1}|^2-|U_{\t2}|^2) \cong - 0.016$, 
$|U_{\t1}^*\, U_{\t2}| \cong 0.27$ and 
the ``high energy'' term will be strongly suppressed, 
being typically by more than an order of magnitude 
smaller (in absolute value) than the ``mixed'' term.
Actually, as we are going to show, in a large 
region of the corresponding parameter space we 
have $|C_{\rm MIX}| > |C_{\rm HE}|$, and therefore 
$|A^{\rm IH}_{\rm MIX}| >  |A^{\rm IH}_{\rm HE}|$.
In certain cases, as the one indicated above, 
one can even get $|C_{\rm MIX}| \gg |C_{\rm HE}|$. 
In these cases the ``high energy'' term 
plays essentially no role in the 
generation of the baryon asymmetry.

  To be concrete, consider the dependence of 
the ``high energy'' and the ``mixed'' terms on
$|R_{12}|$ in the case when
$|R_{12}| \leq |R_{11}| \cong 1$. 
Let us set first $s_{13} = 0$  
and let us work with 
$\sin 2\tilde{\f}_{11} < 0$ and 
\footnote{One can use the following 
property of the baryon asymmetry, 
$Y_B(\tilde{\f}_{11},\tilde{\f}_{12};\a_{21}) = 
- Y_B(-\tilde{\f}_{11},-\tilde{\f}_{12};4\pi -\a_{21})$, 
to obtain results for $\sin 2\tilde{\f}_{11} > 0$
and $\sin 2\tilde{\f}_{12} <0$.}
$\sin 2\tilde{\f}_{12} >0$.
In the region of $|R_{12}| \ltap 0.5$, 
as can be easily verified, we have:
$|\sin 2\tilde{\f}_{11}| \cong |R_{12}|^2$ and therefore 
$|C_{\rm HE}|\propto G_{11}|\sin 2\tilde{\f}_{11}| \propto |R_{12}|^2$.
At the same time
$\tilde{\f}_{12} \cong \pi/4$ and, correspondingly,
for, e.g. $\a_{21}/2 = \pi/4$ we get:
$|C_{\rm MIX}| \propto 2G_{12} |\sin(\tilde{\f}_{11}+
\tilde{\f}_{12})\cos\,\a_{21}/2| \propto |R_{12}|$.
Obviously, in the case under discussion 
one finds $|C_{\rm MIX}| > |C_{\rm HE}|$ and  
the ``mixed'' term dominates over the ``high energy'' term,
$|A^{\rm IH}_{\rm MIX}| >  |A^{\rm IH}_{\rm HE}|$.
This interesting possibility
is illustrated in Fig. \ref{IH_YB_a21_0.5a}. 
We note that $C_{\rm MIX}$ and $C_{\rm HE}$
have opposite signs. 
The details of the behavior of 
$|A^{\rm IH}_{\rm MIX}|$, $|A^{\rm IH}_{\rm HE}|$ and 
of $|Y_B|$ as functions of $|R_{12}|$, seen in the figure, 
can be understood following the behavior of 
$|\sin(\tilde{\f}_{11}+ \tilde{\f}_{12})|$, 
$|\sin 2\tilde{\f}_{11}|$, $\mtil_\t$, 
$\mtil_2$ and correspondingly of
$(\y (0.66\mtil_{\t}) - \y (0.71\mtil_2))$.
For example, the dominating maximum of 
$|A^{\rm IH}_{\rm MIX}|$ at $|R_{12}|\cong 0.4$
corresponds essentially to a maximum of 
$(\y (0.66\mtil_{\t}) - \y (0.71\mtil_2))$ at 
$\mtil_\t \cong 1.2\times 10^{-3}$ eV.
For $|R_{12}| \ltap 0.5$ and $\a_{21}/2 = \pi/4$
we have $\cos(\tilde{\f}_{11}-\tilde{\f}_{12} + 
\a_{21}/2) \cong 1$ and for each given $|R_{12}|$ 
the wash-out mass parameter $\mtil_\t$ has 
its minimal value (see eq. (\ref{mtIHMaj1})).  
As $|R_{12}|$ increases 
starting from the value of 0.5, 
$|\sin 2\tilde{\f}_{11}|$ also increases while 
$|\sin(\tilde{\f}_{11}+ \tilde{\f}_{12})|$ and 
$(\y (0.66\mtil_{\t}) - \y (0.71\mtil_2))$ 
decrease. At $|R_{12}| \cong 0.65$,
$|(\y (0.66\mtil_{\t}) - \y (0.71\mtil_2))|$
goes through a minimum associated 
with $\mtil_{\t}$ having a very small value,
$\mtil_{\t}\cong 0$. At this minimum 
both $|A^{\rm IH}_{\rm MIX}|$ and $|A^{\rm IH}_{\rm HE}|$
are similar in magnitude and
have relatively small
values (Fig. \ref{IH_YB_a21_0.5a}).
Correspondingly, $|Y_B|$ is strongly suppressed
and it is impossible to have successful leptogenesis.
Actually, at $|R_{12}| \cong 0.7$ one finds
$Y_B \cong 0$. At this value of $|R_{12}|$ the 
asymmetry $Y_B$ changes sign. 
As $|R_{12}|$ increases beyond 0.7,
$|\sin 2\tilde{\f}_{11}|$ and 
$(\y (0.66\mtil_{\t}) - \y (0.71\mtil_2))$
continue to increase, while 
$|\sin(\tilde{\f}_{11}+ \tilde{\f}_{12})|$
continues to decrease. As a consequence,
at $|R_{12}| > 0.7$
the ``high energy'' term is larger than the 
``mixed'' term (Fig. \ref{IH_YB_a21_0.5a}) 
and the latter partially compensates the
contribution of the former 
in $Y_B$. At $|R_{12}| \cong 0.85$, 
$(\y (0.66\mtil_{\t}) - \y (0.71\mtil_2))$
has a second maximum. Now  
$|A^{\rm IH}_{\rm HE}|$ dominates over 
$|A^{\rm IH}_{\rm MIX}|$ in $|Y_B|$.

 Obviously, the results we get depend 
critically on the Majorana phase $\a_{21}$. 
If $\a_{21} = 3\pi/2$, for instance, 
$|Y_B|$ will be strongly suppressed 
due to the factor 
$(\y (0.66\mtil_{\t}) - \y (0.71\mtil_2))$
(strong wash-out regime).

   The possibility of strong suppression of the 
``high energy'' term for $s_{13} = 0.2$ and $\delta = \pi$ 
is illustrated for $|R_{12}| \leq |R_{11}| \cong 1$
in Fig. \ref{IH_YB_a21_0.5a_delta_pi}. As the figure shows, 
in this case we have $|A^{\rm IH}_{\rm MIX}| 
\gg  |A^{\rm IH}_{\rm HE}|$ and the 
magnitude of the baryon asymmetry 
$|Y_B|$ is determined entirely by the 
``mixed'' term. A more detailed investigation of
this interesting possibility is presented in 
\cite{Molinaro:2008cw}~
\footnote{ The main differences between the analyses of the case of
IH spectrum performed in the present article 
and in \cite{Molinaro:2008cw} are the following.
Here we have investigated the case
of $R_{13} = 0$, compatible with the hypothesis of
decoupling of the heaviest RH Majorana neutrino $N_3$.
In \cite{Molinaro:2008cw} we have analysed the more
general case of non-zero $R_{13}$ and real $R_{13}^2$.
In the present paper we are interested in
the general interplay between the contributions to the
baryon asymmetry due to the ``high energy'' CP violating
and that due to the ``low energy'' CP violating phases 
in the neutrino mixing matrix, while 
in  \cite{Molinaro:2008cw} we have concentrated only 
on the cases in which  the contribution to the
baryon asymmetry due to the
``high energy'' CP violating phases (the $R$-phases)
is subdominant or strongly suppressed.
Finally, in \cite{Molinaro:2008cw} we give a very detailed
description of the regions of the leptogenesis parameter
space, in which  the contribution to the
baryon asymmetry due to the ``high energy''
CP violating phases is subdominant or
strongly suppressed. This was done both in the cases
$R_{13} = 0$ and of non-zero real $R_{13}^2$.}.
Obviously, the same conclusion 
is valid also for $|R_{11}| > 1$, say $|R_{11}| = 1.2$, 
if $s_{13} = 0.2$ and $\delta = \pi$.

 Finally, in Fig. \ref{IH_YB_a21_0.5a_s13_0_R11_1.2} 
we have presented graphically 
the results of a similar analysis performed 
for $|R_{11}| = 1.2$, $s_{13} = 0$, 
and $\a_{21}/2 = \pi/4$. The ``mixed'' term  
$|A^{\rm IH}_{\rm MIX}|$
in this case can be larger than, or comparable to, 
the ``high energy'' term $|A^{\rm IH}_{\rm HE}|$
in a rather large interval of values of $|R_{12}|$ 
which includes its minimal value, 
$|R_{12}|^2 \sim |R_{11}|^2 - 1$,
and in the ``vicinity'' of its maximal value, 
$|R_{12}|^2 \sim |R_{11}|^2 + 1$.
We can have a complete cancellation between 
the contributions of the 
``mixed'' and ``high energy'' terms 
and $Y_B =0$ at certain value of 
$|R_{12}|$ (Fig. \ref{IH_YB_a21_0.5a_s13_0_R11_1.2}), 
although at the $|R_{12}|$ in question 
each of these two terms is sufficiently large 
to account for the observed value of 
the baryon asymmetry if the other term were not 
present.

%
\subsection{CP Violation Due to Dirac Phase in $U_{\rm PMNS}$ 
and  $R-$Phases}
%
%

  This case corresponds to 
$\alpha_{21} = k\pi$, $k=0,1,2,...$.
The expression for the baryon asymmetry 
has the form given by eq. (\ref{YB_3}).
The ``high energy'' term is the same as in the case 
of CP violation due to the Majorana and $R$ phases
and is determined by eq. (\ref{CHE1}). 
The form of the ``mixed'' term, to be denoted 
as $C^{\rm D}_{\rm MIX}$ (and
$A^{\rm IHD}_{\rm MIX}\propto C^{\rm D}_{\rm MIX}$)
depends on whether $\alpha_{21} = (2k+1)\pi$ 
or $\alpha_{21} = 2q\pi$, $k,q=0,1,2,...$:
\begin{eqnarray}
C^{\rm D1}_{\rm MIX} \cong (-1)^k\,2\,G_{12}\, 
\sin(\tilde{\f}_{11}+\tilde{\f}_{12})\,c_{23}
\,s_{23}\,s_{13}\sin\delta\,,~~\alpha_{21} = (2k+1)\pi\,,
\label{CMIXD1}
\end{eqnarray}
%
\begin{eqnarray}
C^{\rm D2}_{\rm MIX} \cong (-1)^{q+1}\,2G_{12}
\sin(\tilde{\f}_{11}+\tilde{\f}_{12})\,
\left [c_{12}s_{12}s^2_{23} + 
s_{13}\left (s^2_{12} -
c^2_{12}\right )c_{23}s_{23}\cos\delta \right ],
\alpha_{21} = 2q\pi\,.
\label{CMIXD2}
\end{eqnarray}
%
Using the experimentally determined values of 
$s^2_{23}$ and $s^2_{12}$, and the upper limit
$s^2_{13} < 0.05$, it is easy to convince 
oneself that the term involving the Dirac phase 
in  $C^{\rm D2}_{\rm MIX}$, eq. (\ref{CMIXD2}),
gives always a subdominant contribution
to $C^{\rm D2}_{\rm MIX}$. As a consequence,
$C^{\rm D2}_{\rm MIX}$ exhibits very weak 
dependence on $\delta$ even for values of 
$s_{13}$ close to the existing upper limit.
The dominant CP-violating term is due to the 
$R$-matrix. Therefore we will discuss 
only the case $\alpha_{21} = (2k+1)\pi$
in what follows. 

 For arbitrary $\alpha_{21}$,
the wash-out mass parameter $\mtil_{\t}$ reads:
\begin{eqnarray}
\mtil_\t&=&\sqrt{|\dma|}\left | 
\left ( c_{12}|R_{12}| - 
s_{12}|R_{11}|e^{i(\tilde{\f}_{11}-\tilde{\f}_{12} + 
\frac{\a_{21}}{2})}\right )s_{23}\right. \nonumber\\\\ 
&+& 
\left. s_{13}c_{23}e^{-i\delta}
\left (s_{12} |R_{12}| + 
c_{12} |R_{11}|e^{i(\tilde{\f}_{11}-\tilde{\f}_{12} + 
\frac{\a_{21}}{2})}\right )\right |^2.\nonumber
\label{mtIHD1}
\end{eqnarray}
%
The maximum of the ``mixed'' term
$C^{\rm D1}_{\rm MIX}$ with respect to $\delta$
occurs obviously for $\delta = (2k'+1)\pi/2$, $k'=0,1,2,...$.
A more detailed analysis shows that 
for, e.g. $\alpha_{21} = \pi$, 
the absolute maximum of the efficiency term 
$(\y (0.66\mtil_{\t}) - \y (0.71\mtil_2))$
is reached in the case of $\delta \cong 3\pi/2$. 
The interplay between the 
CP-violation due to the 
Dirac Phase in $U$ and  $R-$phases
for $\alpha_{21} = \pi$ and $\delta \cong 3\pi/2$
is illustrated in Figs. \ref{IHD1} and \ref{IHD2}.
In Fig. \ref{IHD1} (Fig. \ref{IHD2}) we show the dependence 
of $|Y^0_B A^{\rm IH}_{\rm HE}|$,
$|Y^0_B A^{\rm IHD1}_{\rm MIX}|$ and of $|Y_B|$  
on $|R_{12}|$ in the case of $|R_{11}|=1$
($|R_{11}|=1.2$). As Figs. \ref{IHD1} and \ref{IHD2} illustrate,
there are substantial ranges of values of 
the relevant parameters for which 
the ``mixed'' term is of the order of, 
or exceeds, the ``high energy'' term.
Typically the contributions of these two 
terms to $Y_B$ have opposite signs and 
tend to compensate each other.
In certain points of the parameter space 
the compensation can be complete
and we can have $Y_B = 0$, 
although both $|Y^0_B A^{\rm IH}_{\rm HE}|$
and $|Y^0_B A^{\rm IHD1}_{\rm MIX}|$
can have relatively large values 
(see Figs. \ref{IHD1} and \ref{IHD2}).
In Fig. \ref{IHD3} 
we show the dependence of the baryon 
asymmetry on the Dirac phase $\delta$
for $\alpha_{21} = \pi$, $s_{13} = 0.2$
and one set of values of 
$|R_{11}|$ and $|R_{12}|$, 
at which the ``mixed'' term dominates in $|Y_B|$ 
when $\delta=3\pi/2$. As the figure indicates, 
$Y_B$ can exhibit very strong dependence 
on the Dirac phase 
$\delta$ when the Majorana phase 
$\alpha_{21}$ takes the CP-conserving 
value $\alpha_{21} = \pi~(3\pi)$.

%
\section{Conclusions}
%
%

 In the present article we have investigated 
the interplay in ``flavoured'' leptogenesis 
between the ``low energy'' CP-violation, 
originating from the 
PMNS  neutrino mixing matrix $U$
and the ``high energy'' CP-violation
which can be present in the matrix of 
neutrino Yukawa couplings, 
$\lambda$, and can manifest itself 
only in ``high'' energy scale phenomena.
We worked within the simplest type I see-saw 
theory with three heavy Majorana
neutrinos $N_j$, $j=1,2,3$, having a 
hierarchical mass spectrum with
masses $M_1 \ll M_{2,3}$.  
In the basis employed by us,
the matrix of neutrino Yukawa 
couplings $\lambda$ is the only source  
of CP-violation in the lepton sector.
In this analysis we used the ``orthogonal'' 
parametrisation of $\lambda$,
involving a complex orthogonal matrix $R$, $R^TR =
RR^T = {\bf 1}$: 
$\lambda \propto \sqrt{M}R \sqrt{m} U^{\dagger}$, 
where $M$ and $m$ are diagonal matrices formed by
the masses $M_j > 0$ and $m_k \geq 0$ of $N_j$ 
and of the light Majorana neutrinos $\nu_k$, $j,k=1,2,3$, 
This parametrisation proved rather convenient in 
the analysis performed in \cite{PPRio106},
which showed that the CP-violation
necessary for a successful leptogenesis 
could be provided exclusively by the 
Majorana and/or Dirac physical 
phases in the neutrino mixing matrix $U$. 
It permitted to investigate also the 
combined effect of the 
CP-violation due to the neutrino 
mixing matrix $U$ and the CP-violation 
due to the matrix $R$ 
in the generation of the baryon asymmetry in 
``flavoured'' leptogenesis. 
Throughout this study we used the terms 
``low energy'' and ``high energy'' 
for the CP-violation originating 
respectively from the matrices $U$ and $R$.
The matrix $R$, as is 
well-known, does not affect the ``low'' 
energy neutrino mixing phenomenology. 
The two matrices $U$ and $R$
are, in general, independent. 
The source of the requisite CP-violation 
in ``flavoured'' leptogenesis can, in principle, be    
the matrix $R$, the PMNS matrix $U$, 
or both $R$ and $U$. 
If the matrix $R$ satisfies the general 
CP-invariance constraints (having real or purely 
imaginary elements \cite{PPRio106}),
while the PMNS matrix $U$
does not satisfy these constraints, 
we consider the CP-violation as originating 
from the neutrino mixing matrix $U$, i.e. 
from the Dirac and/or Majorana phases in $U$.
This case has been studied in detail in 
\cite{PPRio106}. If, however, the Dirac 
and Majorana phases in $U$ 
take CP conserving values, while the matrix $R$, 
and the Yukawa couplings $\lambda$ 
do not satisfy the constraints following from 
the requirement of CP-invariance, 
the CP-violation will manifest itself 
only in ``high'' energy phenomena 
(like, e.g. leptogenesis) and will be due 
to the matrix $R$. As is well-known, 
one can have successful leptogenesis 
in this case as well.
When neither $U$ nor $R$ satisfy 
the CP-invariance conditions,
both $U$ and $R$ will be 
sources of CP-violation effects 
at ``high'' energies .
In the present work we were primarily 
interested in this last possibility.

   In the case of hierarchical 
heavy Majorana neutrinos $N_{1,2,3}$, $M_1
\ll M_{2} \ll M_{3}$, 
the generated baryon asymmetry $Y_B$ depends
(linearly) on the mass of 
the lightest RH Majorana neutrino
$N_1$, $M_1$, and on the elements $R_{1j}$
of the matrix $R$, $j=1,2,3$, 
present in the neutrino Yukawa 
couplings of $N_1$.
The baryon asymmetry was assumed to be produced 
in the ``two-flavour regime'' of leptogenesis, 
which is realised for $M_1 \ltap 10^{12}$ GeV. 
We have considered two types of 
light neutrino mass spectrum allowed 
by the existing data (see, e.g. \cite{STPNu04}), namely,
the normal hierarchical (NH), $m_1 \ll m_2 < m_3$, and 
the inverted hierarchical (IH), $m_3 \ll m_1 < m_2$.
The lightest neutrinos mass in both cases was 
assumed to be negligibly small. Accordingly,  
in the case of NH (IH) spectrum 
we have set also $R_{11} \cong 0$ ($R_{13} \cong 0$), 
which is compatible with the hypothesis of 
decoupling of the heaviest RH Majorana 
neutrino $N_3$ \cite{PRST05,IR041,FGY03}.
Under these conditions the orthogonality 
of the $R$-matrix implies the following constraint 
on the elements $R_{1j}$ of interest:
$R^2_{12} + R^2_{13} = 1$ ($R^2_{11} + R^2_{12} = 1$).
 The phases of $R_{12}$ and $R_{13}$ ($R_{11}$ and $R_{12}$),
$\tilde{\f}_{12}$ and $\tilde{\f}_{13}$ 
($\tilde{\f}_{11}$  and $\tilde{\f}_{12}$), 
are the ``high energy'' CP-violating leptogenesis 
parameters. Using the orthogonality condition, 
$\tilde{\f}_{12}$ and $\tilde{\f}_{13}$ 
($\tilde{\f}_{11}$ and $\tilde{\f}_{12}$) 
can be expressed in terms of 
$|R_{12}|$ and $|R_{13}|$ ($|R_{11}|$ and $|R_{12}|$).
Thus, for fixed $M_1$ and given neutrino oscillation 
parameters $|\dma|$, $\dmsol$, $\sin^2\theta_{23}$, 
$\sin^2\theta_{12}$ and $\sin^2\theta_{13}$,
the baryon asymmetry in the cases we have studied 
depends, in general, on two ``high energy'' 
CP-violating ($R$-) phases and on
one Majorana and one Dirac 
``low energy'' CP-violating 
phases of the neutrino 
mixing matrix $U$.
Under the conditions considered 
the Majorana phase which enters into 
the expression for $Y_B$ corresponding 
to NH (IH) spectrum is 
$\alpha_{32} \equiv \alpha_{31} - \alpha_{21}$
($\alpha_{21}$).

  Analyzing the possibility of NH spectrum 
and CP violation due to the Majorana 
phase $\alpha_{32}$ and the $R$-phases 
$\tilde{\f}_{12}$ and $\tilde{\f}_{13}$ 
(subsection 3.1), we have found that 
there exists a relatively large region 
of the relevant parameter space 
in which the predicted value of 
the baryon asymmetry exhibits a 
strong dependence on 
the Majorana phase 
$\alpha_{32}$ provided the latter lies 
in the interval $0< \alpha_{32} < \pi$
(if $\sin 2\tilde{\f}_{12} < 0$, $\sin 2\tilde{\f}_{13}> 0$), 
or  $3\pi < \alpha_{32} < 4\pi$
(when $\sin 2\tilde{\f}_{12} > 0$, $\sin 2\tilde{\f}_{13}< 0$).
The regions typically correspond 
to $0.05 \ltap |R_{13}| \ltap 0.5$, 
$|R_{13}| < |R_{12}| \leq 1$, 
and to $|R_{12}| > 1$, 
$|R_{13}|^2 \sim |R_{12}|^2 - 1$.
Depending on the value of  $\alpha_{32}$, 
we can have, e.g.
either $|Y_B| \ll 8.6\times 10^{-11}$ or 
$Y_B$ compatible with the observations
in the indicated regions.
The effects of the ``low energy'' CP violation 
due to $\alpha_{32}$ can be non-negligible 
in leptogenesis also for 
$0.5 \leq |R_{13}| \leq |R_{12}| \leq 1$ 
(Figs. 1, 2 and 6).
In the regions of the parameter space 
where the Majorana phase effects are 
significant, the contributions to 
$Y_B$ due to the ``high energy'' 
CP-violation and that involving 
the ``low energy'' CP-violating phase 
$\alpha_{32}$ typically have opposite 
signs and tend to compensate each other.
This mutual compensation 
can be complete and we can have 
$Y_B = 0$ for certain values of the 
relevant parameters, in spite of the fact 
that each of the two contributions can be 
sufficiently large to account 
by itself for the observed value of 
$Y_B$. We have found also that in the 
regions of values of the parameters 
for which there is a significant 
interplay between the ``high energy''
and the ``low energy'' CP-violation, 
the predicted value of $|Y_B|$ 
can exhibit strong dependence 
i) on the atmospheric neutrino 
mixing parameter $\sin^2\theta_{23}$ 
when the latter is varied 
in the range (0.36 - 0.64) allowed by the data, 
and, ii) on whether the 
Dirac phase $\delta = 0~{\rm or}~\pi$, 
if the value of $\sin^2\theta_{13}$ 
is sufficiently large (Figs. 3 - 5).    

  If in the case of NH spectrum the Majorana phase 
$\alpha_{32}$ has a CP-conserving value, 
$\alpha_{32} = 2\pi k$, $k=0,1,2,...$, 
the ``low energy'' CP-violation is due 
only to the Dirac phase $\delta$ in the 
PMNS matrix $U$ (subsection 3.2).  
A region in the corresponding 
parameter space where we have 
noticeable CP-violation effects
due to $\delta$ and successful leptogenesis 
still exists, but is very limited
(Fig. 7). This is essentially 
a consequence of the suppression by the 
factor $\sin\theta_{13} < 0.22$ of the 
CP-violation effects associated with $\delta$.
For CP-violating values of the
``high energy'' phases $\tilde{\f}_{12}$ 
and $\tilde{\f}_{13}$ such that 
$\sin(\tilde{\f}_{12} + \tilde{\f}_{13})$
is significantly different from zero,
the contribution of the term in 
$Y_B$ involving the Dirac phase 
is actually further suppressed.

  We obtained very different results for  
IH neutrino mass spectrum (Section 4).
In this case there are large regions of values 
of the corresponding parameters, for which 
the contribution to $Y_B$ due to
the ``low energy'' CP-violating Majorana phase 
$\alpha_{21}$, or Dirac phase $\delta$ 
(for $\alpha_{21} = (2k + 1)\pi$), 
is comparable in magnitude, or exceeds, 
the purely ``high energy'' contribution 
in $Y_B$, originating from 
CP-violation generated by the $R$-matrix 
(Figs. 8 - 13). Moreover, in certain significant 
subregions of the indicated regions, 
the contribution to $Y_B$ due to 
the ``high energy'' CP-violation is 
subdominant. We have found also that for 
$(-\sin\theta_{13}\cos \delta)\gtap 0.1$,
the ``high energy'' term in $Y_B$ 
is strongly suppressed by the factor 
$(|U_{\t1}|^2-|U_{\t2}|^2)$. 
The ``high energy'' phases 
$\tilde{\f}_{11}$ and $\tilde{\f}_{12}$ 
in this case can have 
large CP-violating values.
Nevertheless, if the 
indicated inequality is fulfilled,
the purely ``high energy'' contribution to $Y_B$ 
due to the CP-violating $R$-phases 
would play practically no role 
in the generation of baryon asymmetry 
compatible with the observations. 
One would have successful leptogenesis in this case 
only if the requisite CP-violation 
is provided by the Majorana and/or Dirac phases 
in the neutrino mixing matrix.

  The results obtained in this study show that
the CP-violation due to the ``low energy'' 
Majorana and Dirac phases 
in the neutrino mixing matrix 
can play a significant role in the 
production of baryon asymmetry 
compatible with the observation in 
``flavoured'' leptogenesis 
even in the presence of  ``high energy'' 
CP-violation  generated by additional physical 
phases in the matrix of neutrino 
Yukawa couplings, e.g. by CP-violating phases in 
the complex orthogonal matrix $R$
appearing in the ``orthogonal parametrisation'' 
of neutrino Yukawa couplings.

{\bf Note Added.} In the article \cite{SDavid08} 
which appeared approximately 3 months after our paper, 
negative results regarding the possible 
effects of ``low energy'' CP violation in 
``flavoured'' leptogenesis with
hierarchical heavy (RH) and light Majorana 
neutrinos, when the  ``high energy'' 
CP-violation is also present,
were reported. It should be noted  that  
the analysis performed in \cite{SDavid08} differs 
substantially from the analysis performed here.
In \cite{SDavid08} the leptogenesis is 
considered in the framework of the 
SUSY extension of the Standard Model, 
more specifically, in the 
minimal Supergravity (MSUGRA)
scenario with real boundary conditions, 
in which the dynamics responsible for supersymmetry
breaking are flavour blind and all the lepton flavour and 
CP violation is controlled by the neutrino Yukawa couplings. 
The leptogenesis parameter space 
is constrained, in particular,
by requiring that the $\mu \rightarrow e + \gamma$ 
decay rate branching ratio, predicted in this scenario,
satisfies $BR(\mu \rightarrow e + \gamma) \geq 10^{-12}$. 
We work in the simpler non-SUSY version 
of leptogenesis. The difference between our results and 
those found in \cite{SDavid08}, as the authors of 
\cite{SDavid08} also notice, may reflect the difference in the 
priors on the scanned leptogenesis parameters, 
for instance, the range in which 
the lightest RH neutrino mass $M_1$ is varied, 
In any case, the aim of the analysis performed in
\cite{SDavid08} is different from the motivation 
and the resulting conclusions of our paper, as is 
clearly explained in the introduction of \cite{SDavid08}.

\vspace{-0.4cm}
\section{Acknowledgments}
\vspace{-0.3cm}

 This work was supported in part by the INFN 
under the program ``Fisica Astroparticellare'', 
by the Italian MIUR program  
on ``Fundamental Constituents of the Universe''
and by the  European Network of Theoretical 
Astroparticle Physics ILIAS/N6 
(contract RII3-CT-2004-506222). 
S.T.P. acknowledges with gratefulness 
the hospitality and support of 
IPMU, University of Tokyo, 
where part of the work on the 
present article was done. 


\newpage

\begin{figure*}[t!!]
\begin{center}
\vspace{-1cm}
\includegraphics[width=13.5cm,height=9.0cm]{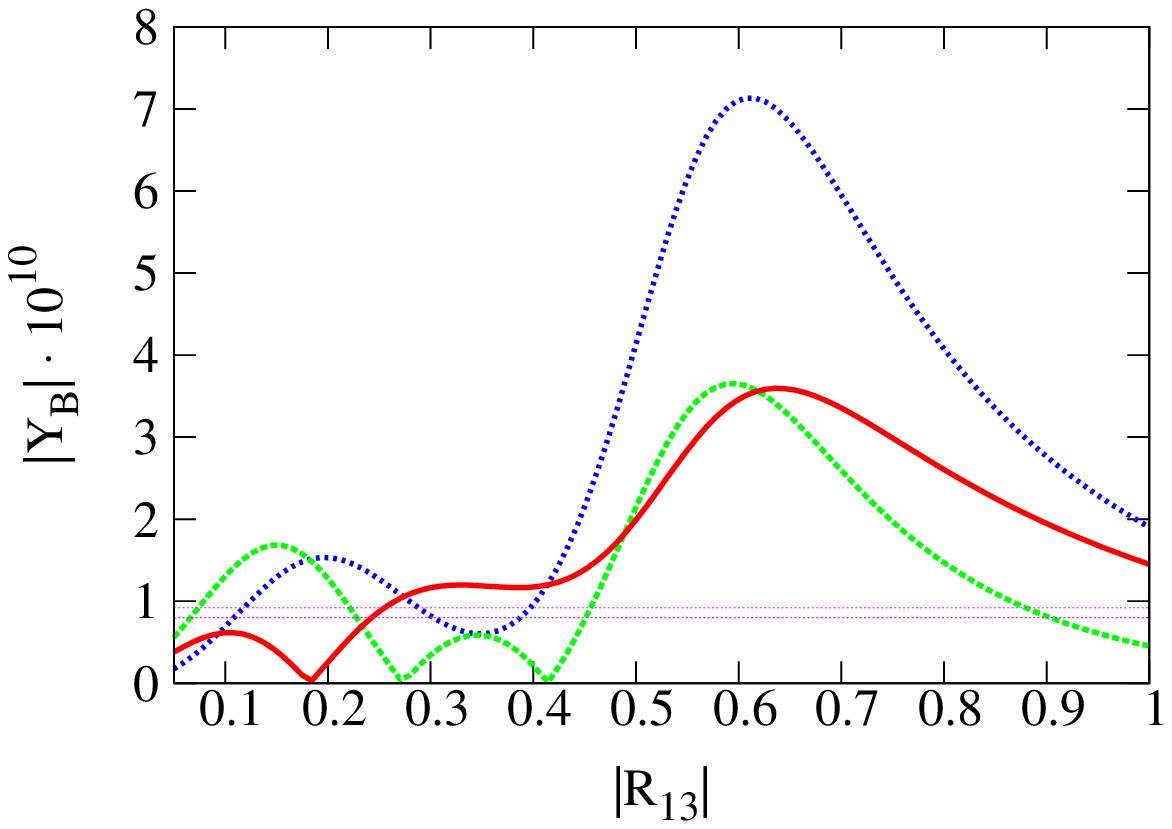} 
\caption{
\label{NH_YB_1a} 
The dependence of the ``high energy'' term 
$|Y^0_B A_{\rm HE}|$ (blue line), the ``mixed'' term
$|Y^0_B A_{\rm MIX}|$ (green line) and of the 
total baryon asymmetry $|Y_B|$ (red line) 
on $|R_{13}|$ in the case of NH spectrum, 
CP-violation due to the Majorana phases in $U$ 
and $R$-phases, $\alpha_{32} = \pi/2$,
$s_{23}^2=0.5$, $s_{13}=0$, $|R_{12}|\cong 1$ and $M_1=10^{11}\GeV$.
}
\end{center}
\end{figure*}

\begin{figure*}[t!!]
\begin{center}
\vspace{-1cm}
\includegraphics[width=13.5cm,height=9.0cm]{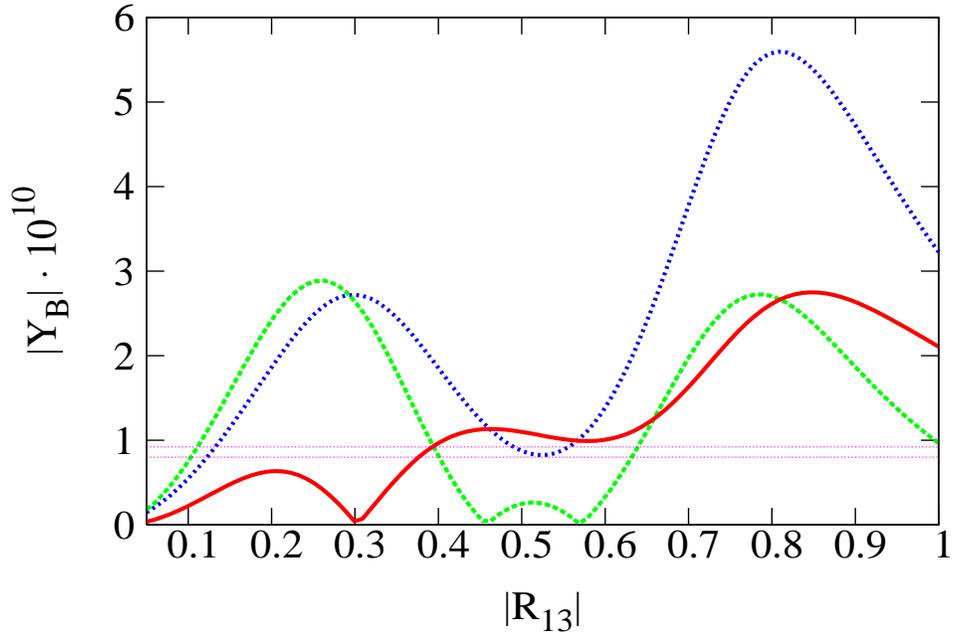} 
\caption{
  \label{NH_YB_1ab} 
The same as in Fig. \ref{NH_YB_1a}, but for 
$s_{23}^2=0.64$, $s_{13}=0.2$ and $\d=0$. 
}
\end{center}
\end{figure*}

\begin{figure*}[t!!]
\begin{center}
\vspace{-1cm}
\includegraphics[width=13.5cm,height=9.0cm]
{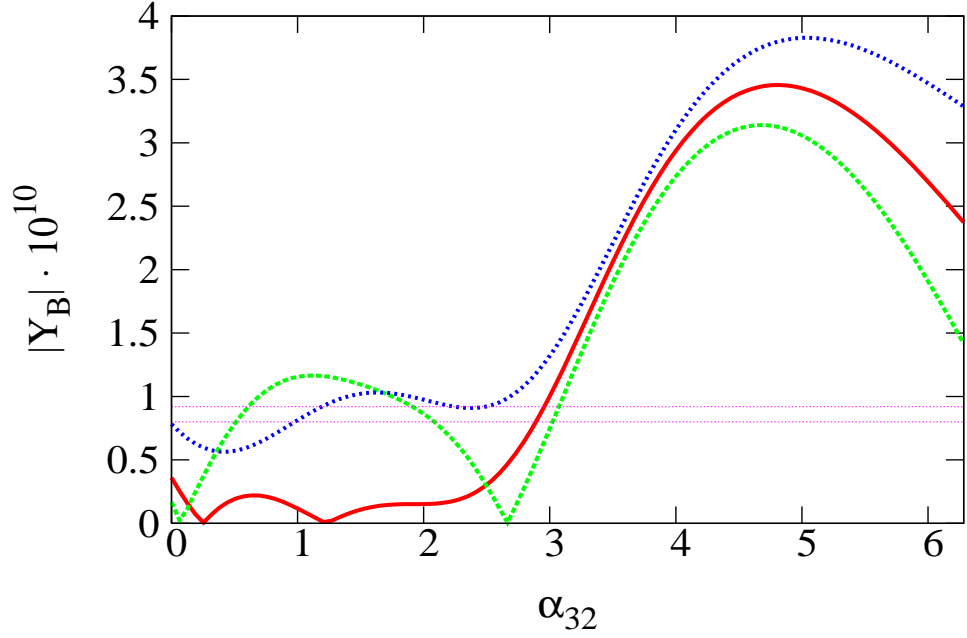}
\caption{
\label{NH_YB_5} 
The dependence of $|Y_B|$ on the Majorana phase 
(difference) $\a_{32}$ in the case of NH spectrum, 
Majorana and $R$ matrix CP-violation, 
$s_{23}^2=0.5$, $M_1=2\times 10^{11}\,\GeV$, 
$R_{12} \cong 1$, $R_{13} = 0.19$,
i) $s_{13}=0$ (red line), 
ii) $s_{13}=0.2$, $\d=0$ (green line), 
iii) $s_{13}=0.2$, $\d=\pi$ (blue line).
}
\end{center}
\end{figure*}

\begin{figure*}[t!!]
\begin{center}
\vspace{-1cm}
\includegraphics[width=13.5cm,height=9.0cm]{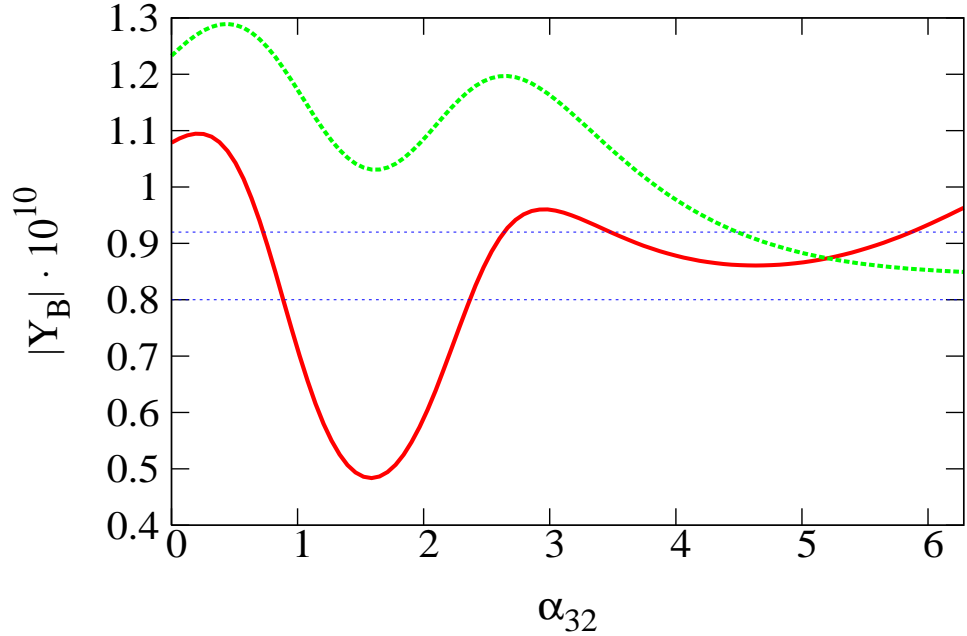}
\caption{
  \label{NH_YB_6} 
The dependence of $|Y_B|$ on $\a_{32}$ in
the case of NH spectrum, Majorana and $R$ matrix CP-violation, 
$|R_{12}| = 1$, $|R_{13}| = 0.51$, 
$M_1= 3.5\times 10^{10}\,\GeV$,
$s_{23}^2=0.5$, $s_{13}=0.2$
and $\d = 0\,(\pi)$ (red (green) line).
}
\end{center}
\end{figure*}

\begin{figure*}[t!!]
\begin{center}
\vspace{-1cm}
\includegraphics[width=13.5cm,height=9.0cm]{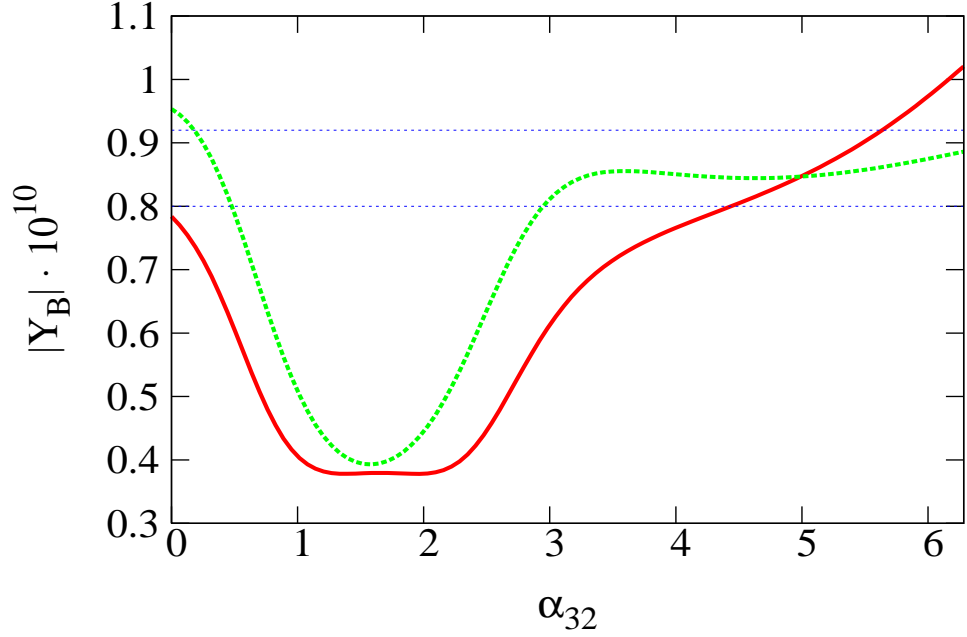}
\caption{
  \label{NH_YB_8} 
The same as in Fig. \ref{NH_YB_6} but for $s_{23}^2=0.64$.
}
\end{center}
\end{figure*}

\begin{figure*}[t!!]
\begin{center}
\vspace{-1cm}
\includegraphics[width=13.5cm,height=9.0cm]{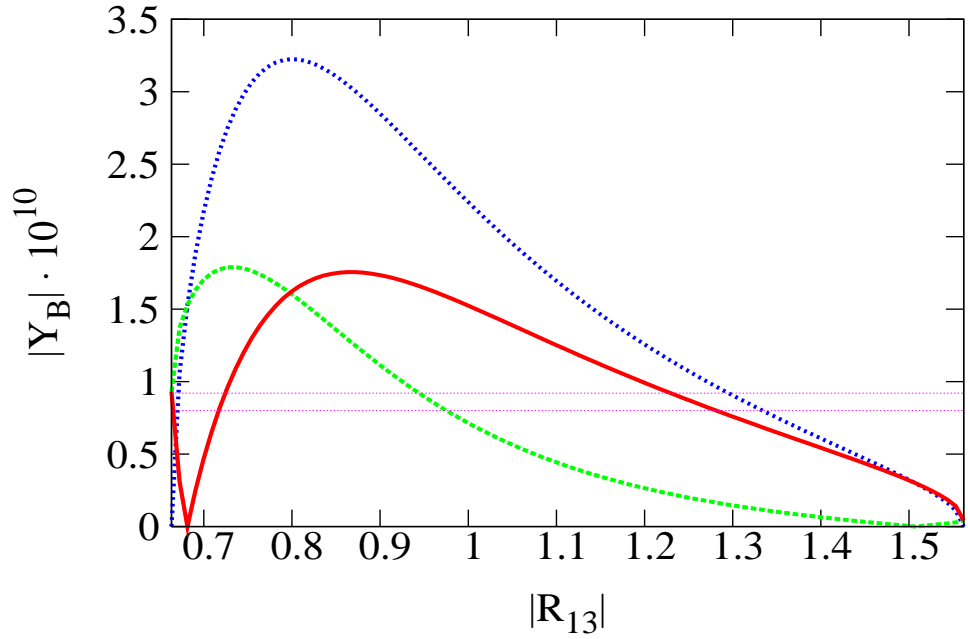} 
\caption{
  \label{NH_YB_1abc}
The dependence of $|Y^0_B A_{\rm HE}|$ (blue line),
$|Y^0_B A_{\rm MIX}|$ (green line) and of $|Y_B|$ (red line) 
on $|R_{13}|$ in the case of NH spectrum, 
Majorana and R-matrix CP-violation,
$|R_{12}|=1.2$, $\alpha_{32}/2 = \pi/4$,
$s_{23}^2=0.5$, $s_{13}=0$  and $M_1=10^{11}\GeV$.
}
\end{center}
\end{figure*}

\begin{figure*}[t!!]
\begin{center}
\vspace{-1.0cm}
\begin{tabular}{cc}
\includegraphics[width=8truecm,height=6.5cm]
{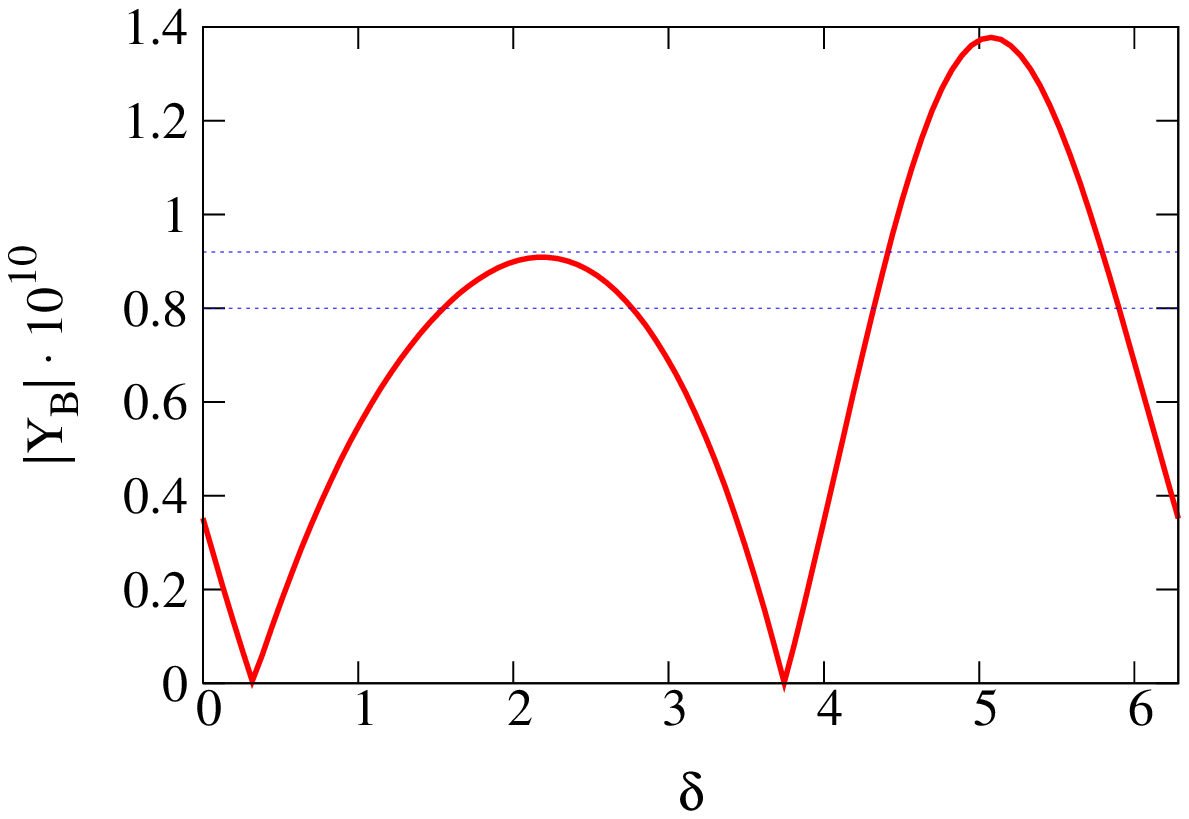}&
\includegraphics[width=8truecm,height=6.5cm]
{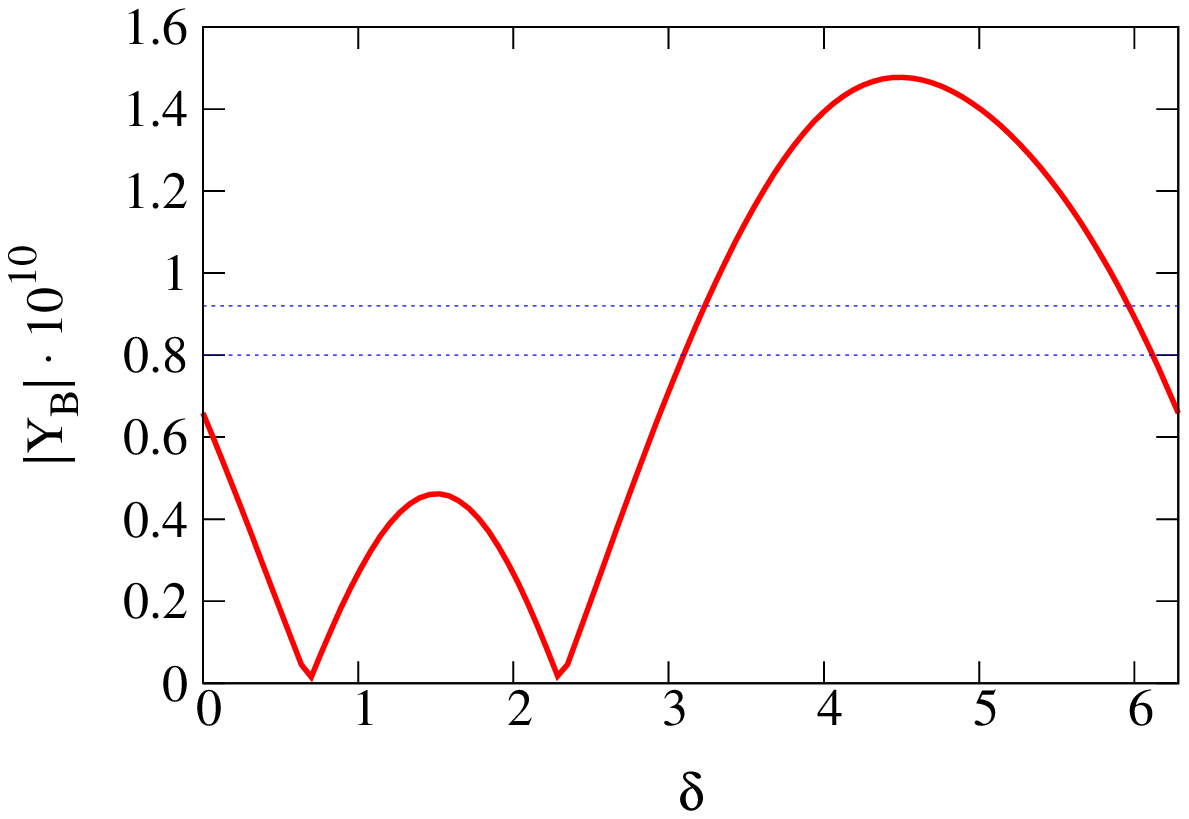}
\end{tabular}
\caption{\label{NH_YB_delta} The dependence of $|Y_B|$ on the 
Dirac phase $\d$ in the case of NH spectrum, Dirac 
and $R$-matrix CP-violation,
$s_{13}=0.2$, 
$R_{12}\cong 1$, $M_1=5\times 10^{11}\,\GeV$ and for 
i) $\alpha_{32} = 0$, 
$|R_{13}|\cong 0.16$ (left panel) and 
ii) $\alpha_{32} = \pi$, 
$|R_{13}|\cong 0.12$ (right panel).}
\end{center}
\end{figure*}


\begin{figure*}[t!!]
\begin{center}
\vspace{-1cm}
\includegraphics[width=13.5cm,height=9cm]
{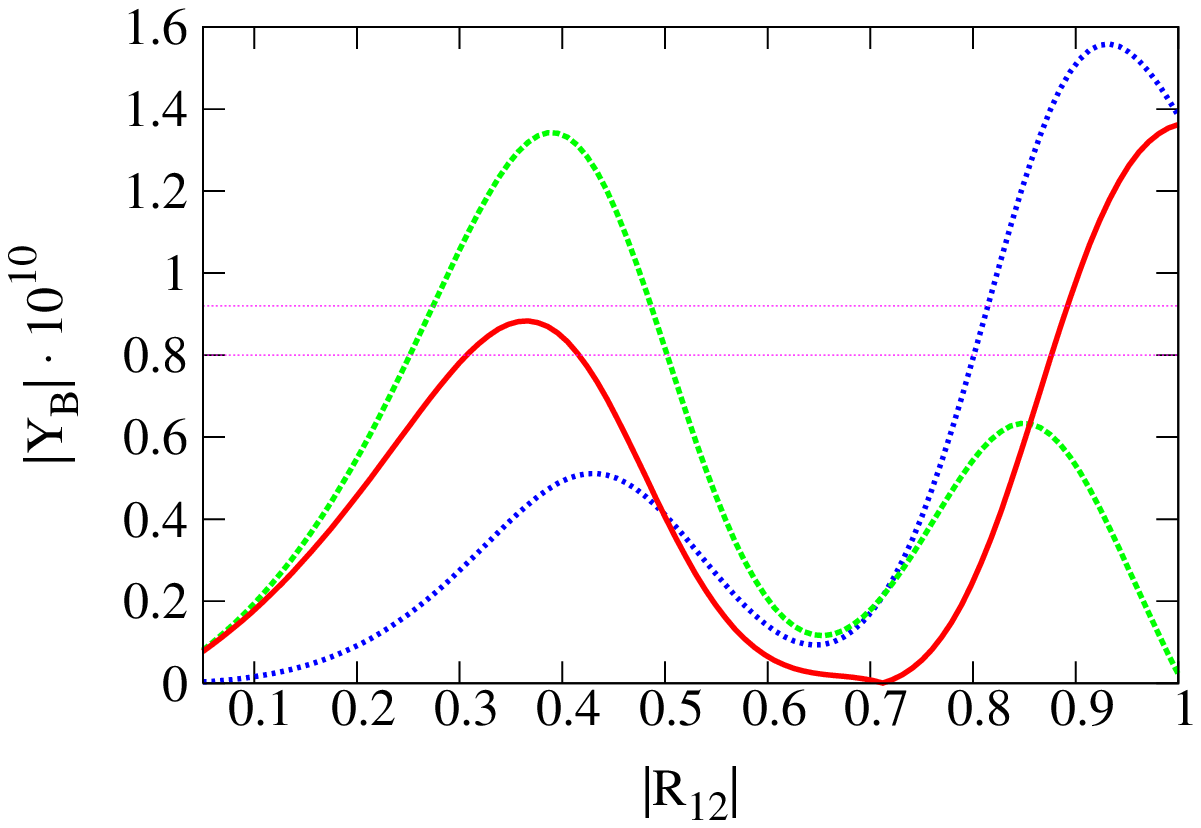}
\caption{
\label{IH_YB_a21_0.5a} 
The dependence of 
the ``high energy'' term 
$|Y^0_B A^{\rm IH}_{\rm HE}|$ (blue line), the ``mixed'' term
$|Y^0_B A^{\rm IH}_{\rm MIX}|$ (green line) and of the 
total baryon asymmetry $|Y_B|$ (red line) 
on $|R_{12}|$ in the case of IH spectrum, 
CP-violation due to the Majorana phase
$\alpha_{21}$ in $U$ and $R$-phases,
for $\alpha_{21} = \pi/2$, $|R_{11}| \cong 1.0$,
$s_{13}=0$ and $M_1 = 10^{11}$ GeV.
}
\end{center}
\end{figure*}

\begin{figure*}[t!!]
\begin{center}
\vspace{-1cm}
\includegraphics[width=13.5cm,height=9cm]
{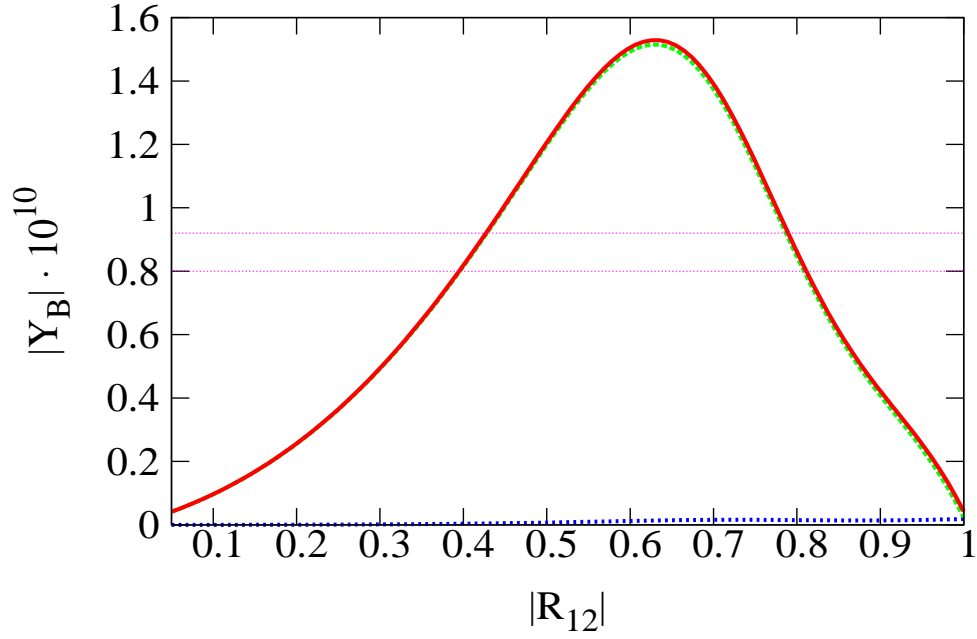}
\caption{
\label{IH_YB_a21_0.5a_delta_pi} 
The same as in Fig. \ref{IH_YB_a21_0.5a}, but for
$s_{13}=0.2$ and $\d=\pi$.
The ``high energy'' term 
$|Y^0_B A^{\rm IH}_{\rm HE}|$ (blue line)
is strongly suppressed.
}
\end{center}
\end{figure*}

\begin{figure*}[t!!]
\begin{center}
\vspace{-1cm}
\includegraphics[width=13.5cm,height=9cm]
{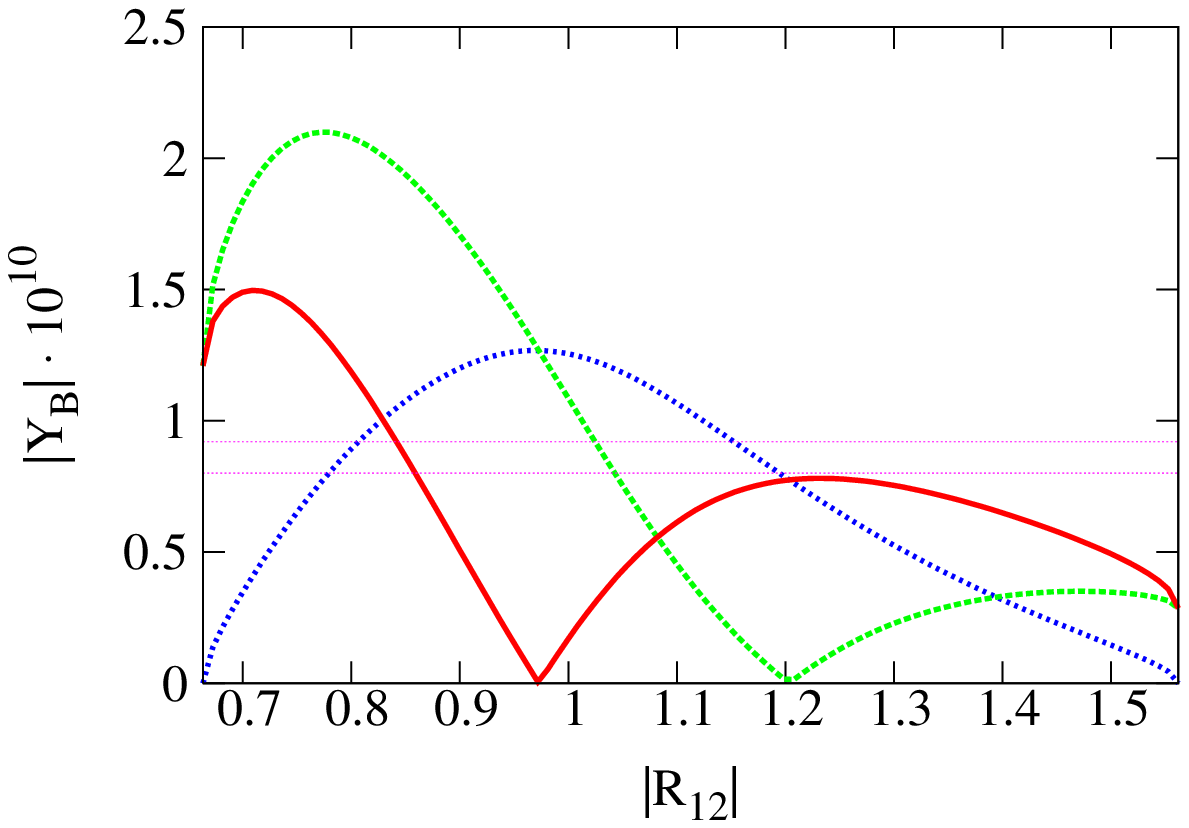}
\caption{
\label{IH_YB_a21_0.5a_s13_0_R11_1.2} 
The dependence of 
the ``high energy'' term 
$|Y^0_B A^{\rm IH}_{\rm HE}|$ (blue line), the ``mixed'' term
$|Y^0_B A^{\rm IH}_{\rm MIX}|$ (green line) and of the 
total baryon asymmetry $|Y_B|$ (red line) 
on $|R_{12}|$ in the case of IH spectrum, 
CP-violation due to the Majorana phase
$\alpha_{21}$ in $U$ and $R$-phases,
for $\alpha_{21} = \pi/2$, $|R_{11}| \cong 1.2$,
$s_{13}=0$ and $M_1 = 10^{11}$ GeV.
}
\end{center}
\end{figure*}

\begin{figure*}[t!!]
\begin{center}
\vspace{-1cm}
\includegraphics[width=13.5cm,height=9cm]
{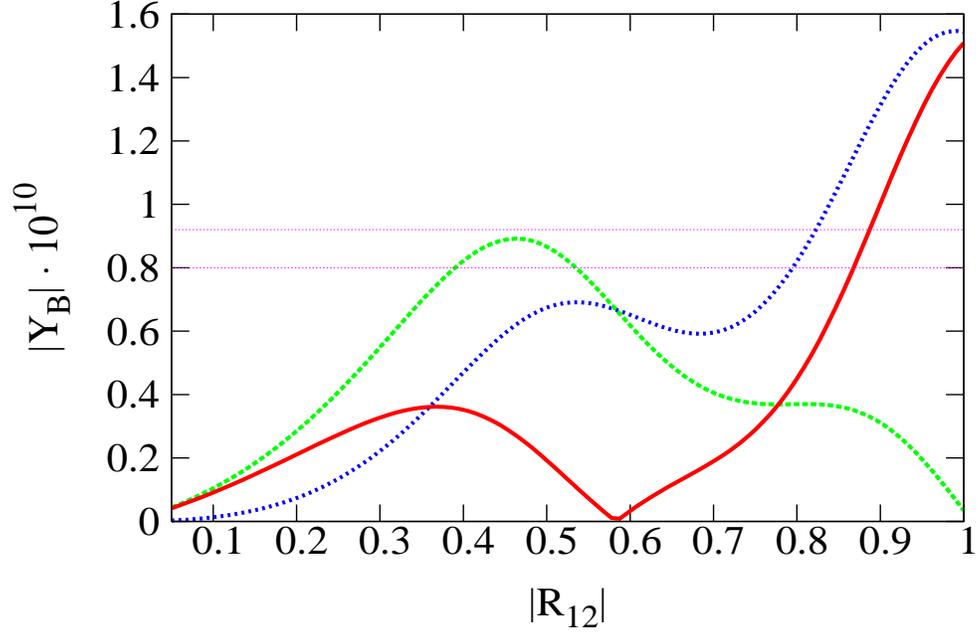}
\caption{
\label{IHD1} 
The dependence of 
the ``high energy'' term 
$|Y^0_B A^{\rm IH}_{\rm HE}|$ (blue line), the ``mixed'' term
$|Y^0_B A^{\rm IH}_{\rm MIX}|$ (green line) and of the 
total baryon asymmetry $|Y_B|$ (red line) 
on $|R_{12}|\leq 1$ in the case of IH spectrum, 
CP-violation due to the Dirac phase
$\delta$ in $U$ and $R$-phases,
$s_{13} = 0.2$, $\delta=3\pi/2$,
$\alpha_{21} = \pi$, $|R_{11}| \cong 1.0$ 
and $M_1 = 10^{11}$ GeV.
}
\end{center}
\end{figure*}

\begin{figure*}[t!!]
\begin{center}
\vspace{-1cm}
\includegraphics[width=13.5cm,height=9cm]
{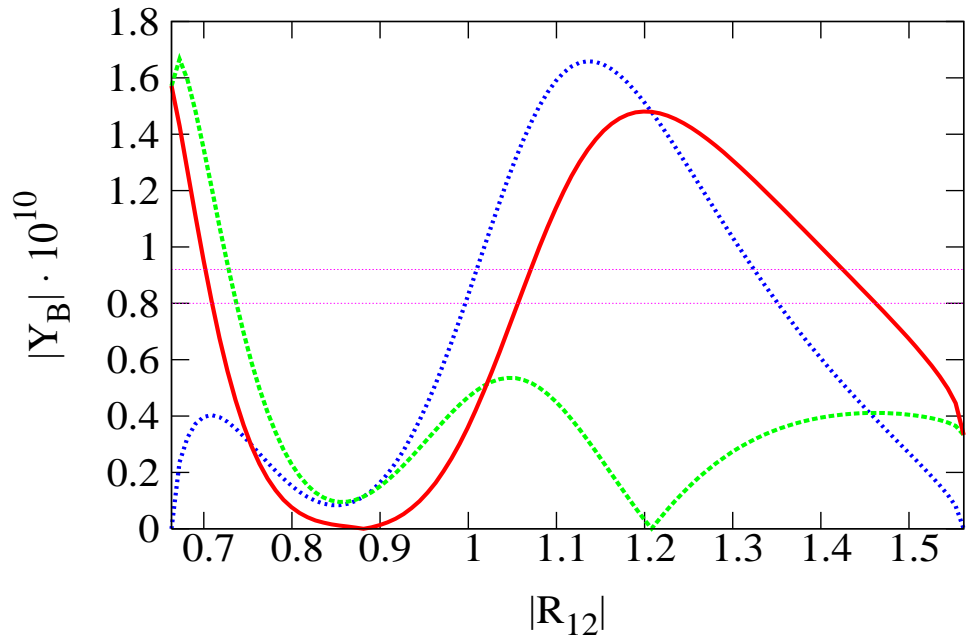}
\caption{
\label{IHD2} 
The same as in Fig. \ref{IHD1} but for 
$|R_{11}| = 1.2$.
}
\end{center}
\end{figure*}

\begin{figure*}[t!!]
\begin{center}
\vspace{-1cm}
\includegraphics[width=13.5cm,height=9cm]
{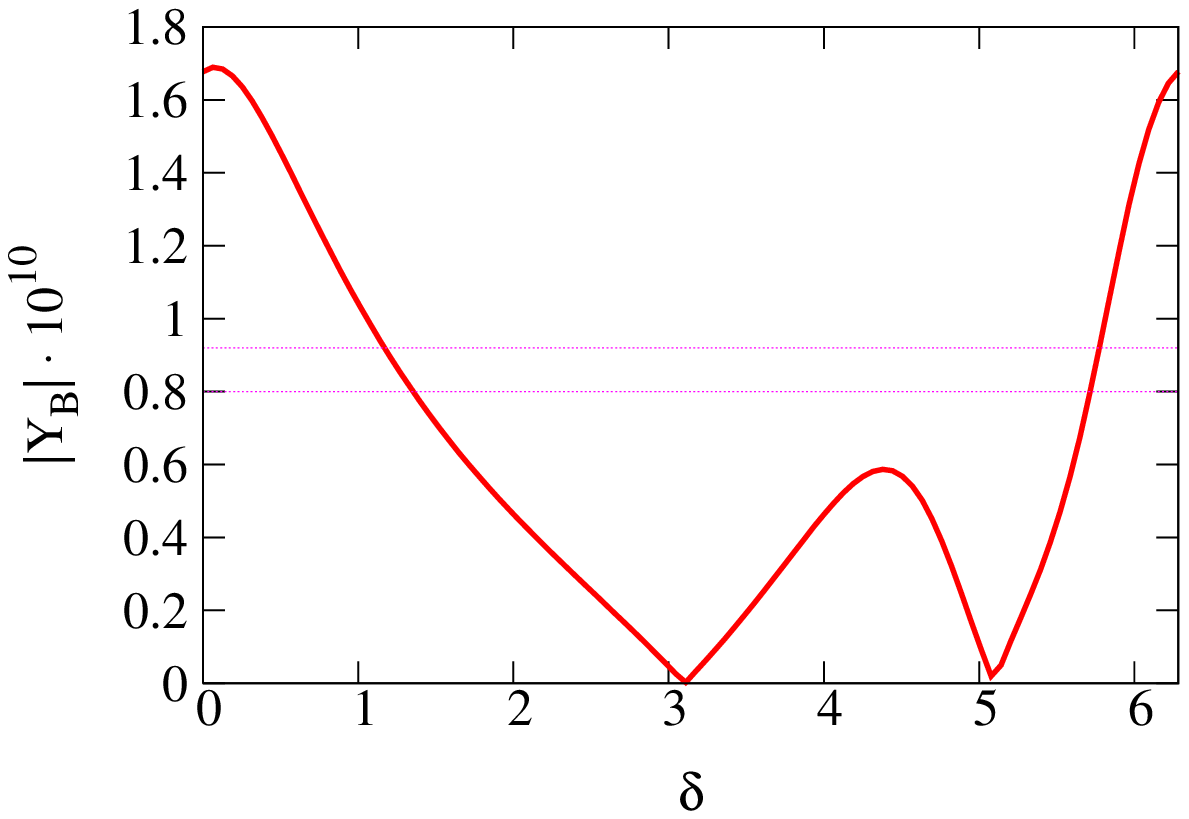}
\caption{
\label{IHD3} 
The dependence of the 
baryon asymmetry $|Y_B|$ 
on the Dirac phase $\delta$ 
in the case of IH spectrum, 
Dirac and $R$ phases CP-violation,
$s_{13} = 0.2$, $\alpha_{21} = \pi$, $|R_{11}| = 1$,
$|R_{12}| = 0.45$ and $M_1 = 1.5\times 10^{11}$ GeV.
}
\end{center}
\end{figure*}


\begin{thebibliography}{99}

\bibitem{FY} M.~Fukugita and T.~Yanagida,
Phys.\ Lett.\  B {\bf  174} (1986) 45.

\bibitem{kuzmin} V.A. Kuzmin, V.A. Rubakov 
and M.E. Shaposhnikov, 
Phys. Lett. B {\bf  155} (1985) 36.

\bibitem{LG1}
W.~Buchm\"uller, P.~Di Bari and M.~Pl\"umacher,
Nucl. Phys. B {\bf 643} (2002) 367;
Annals Phys.\  {\bf 315} (2005) 305.

\bibitem{LG2}
G.~F.~Giudice {\it et al.}, Nucl.\ Phys.\  B {\bf 685} (2004) 89.


\bibitem{others} H.~B.~Nielsen and Y.~Takanishi, 
Phys. Lett. B {\bf 507} (2001) 241; 
W.~Buchm\"uller and D.~Wyler, 
Phys.\ Lett.\ B {\bf 521} (2001) 291; 
J.~Ellis, M.~Raidal and T.~Yanagida, 
Phys. \ Lett.\ B {\bf 546} (2002) 228; 
S.~Davidson and A.~Ibarra, 
Nucl.\ Phys.\ B {\bf 648} (2003) 345;
T.~Endoh, T.~Morozumi and Z.~h.~Xiong,
Prog.\ Theor.\ Phys.\  {\bf 111} (2004) 123.

\bibitem{others2} 
M.~Hirsch, S.~F.~King, 
Phys. Rev. D {\bf 64} (2001) 113005; 
G.C.\ Branco {\it et al.},  
Nucl.\ Phys.\ B {\bf 640} (2002) 202;
J. Ellis and M. Raidal,  Nucl.\ Phys.\ B {\bf 643} (2002) 229;
M.N. Rebelo, Phys. Rev. D {\bf 67} (2003) 013008.

\bibitem{PPR03} S. Pascoli, S.T. Petcov and W. Rodejohann, 
Phys. Rev. D {\bf 68} (2003) 093007.

\bibitem{PRST05}
S.T.~Petcov, W.~Rodejohann, T.~Shindou and Y.~Takanishi,
Nucl.\ Phys.\ B {\bf 739} (2006) 208.

\bibitem{PPRio106} S. Pascoli, S.T. Petcov and A. Riotto, 
Phys. Rev. D {\bf 75} (2007) 083511;
Nucl.\ Phys.\ B {\bf 774} (2007) 1.

\bibitem{Branco06}
S. Antusch, S.F. King and A. Riotto,
JCAP {\bf 0611} (2006) 011;
G.C. Branco {\it et al.},
JHEP {\bf 09} (2007) 004.

\bibitem{BPont57} B.~Pontecorvo, 
                  Zh.\ Eksp.\ Teor.\ Fiz.\ {\bf 33} (1957) 549, 
{\bf 34} (1958) 247 and {\bf 53} (1967) 1717;
Z.~Maki, M.~Nakagawa and S.~Sakata, 
Prog.\ Theor.\ Phys.\  {\bf 28} (1962) 870.
%

\bibitem{Barbieri99}
R.~Barbieri, P.~Creminelli, A.~Strumia and N.~Tetradis,
Nucl. Phys. B {\bf 575} (2000) 61.

\bibitem{Nielsen02}
H.~B.~Nielsen and Y.~Takanishi,
Nucl.\ Phys.\  B {\bf 636} (2002) 305.

\bibitem{davidsonetal} 
A.~Abada {\it et al.},
JCAP {\bf 0604} (2006) 004; 
E.~Nardi, Y.~Nir, E.~Roulet and J.~Racker,
JHEP {\bf 0601} (2006) 164.

\bibitem{davidsonetal2}
A.~Abada {\it et al.},
JHEP {\bf 0609} (2006) 010.


\bibitem{MPST07} 
E. Molinaro, S.T. Petcov, T. Shindou and Y. Takanishi,
{Nucl.\ Phys.} B {\bf 797} (2008) 93. 


\bibitem{SBDibari06}
 S.~Blanchet and P.~Di Bari,
  JCAP {\bf 0703} (2007) 018; 
 G.~C.~Branco, R.~Gonzalez Felipe and F.~R.~Joaquim,
  Phys.\ Lett.\  B {\bf 645} (2007) 432;
 A. Aniusimov, S. Blanchet and 
P.~Di Bari, arXiv:0707.3024 [hep-ph].

\bibitem{DiBGRaff06}
  S.~Blanchet, P.~Di Bari and G.G.~Raffelt,
  JCAP {\bf 0703} (2007) 012.


\bibitem{seesaw}  P. Minkowski,
Phys.\ Lett.\  B {\bf  67} (1977) 421;
M. Gell-Mann, P. Ramond and R. Slansky, 
{\em Proceedings of the Supergravity Stony Brook Workshop}, 
New York 1979,  eds. P. Van Nieuwenhuizen and D. Freedman;
T. Yanagida,  
{\em Proceedings of the Workshop on Unified Theories and Baryon Number in the
Universe},  Tsukuba, Japan 1979, ed.s A. Sawada and A. Sugamoto;
 R. N. Mohapatra and G. Senjanovic, Phys. Rev. Lett. {\bf 44} (1980) 912.

\bibitem{STPNu04} S.T.~Petcov,
{Nucl.\ Phys.\ B (Proc. Suppl.)}
{\bf 143} (2005) 159 (hep-ph/0412410).


\bibitem{MoscowH3Mainz}
V.~Lobashev {\it et al.},  
{Nucl.\ Phys.}  A {\bf 719}(2003) 153c;
%
K.~Eitel {\it et al.}, {Nucl. Phys. B (Proc. 
Suppl.)} {\bf 143} (2005) 197.
%

\bibitem{Hann06}  S.~Hannestad, H.~Tu and Y.Y.Y.~Wong,
  JCAP {\bf 0606} (2006) 025.


\bibitem{Casas01}
J.~A.~Casas and A.~Ibarra,
Nucl. Phys.  B {\bf  618} (2001) 171.


\bibitem{BHP80} S.M. Bilenky, J. Hosek and S.T. Petcov,
             {Phys. Lett.} B {\bf 94} (1980) 495.

\bibitem{Lang87} P. Langacker {\it et al.},  
{Nucl. Phys.} B {\bf 282} (1987) 589.


\bibitem{BiPet87} S.M. Bilenky and S.T. Petcov,
                {Rev. Mod. Phys.} {\bf 59} (1987) 67. 

\bibitem{PKSP3nu88} P.I.~Krastev and S.T.~Petcov,
Phys.\ Lett.\ B {\bf 205} (1988) 84.

\bibitem{Future} 
C.~Albright {\it et al.}, physics/0411123;
Y.~Itow {\it et al.}, hep-ex/0106019;
D.~S.~Ayres {\it et al.}, hep-ex/0503053;
A. Bandyopadhyay {\it et al.}, arXiv:0710.4947.

\bibitem{PSMoriond06} S.T. Petcov and T. Shindou, hep-ph/0605151.
%

\bibitem{BPP1} 
 S.~M.~Bilenky, S.~Pascoli and S.~T.~Petcov,
  Phys.\ Rev.\  D {\bf 64} (2001) 113003.

\bibitem{STPFocusNu04} S.T.~Petcov, New J.\ Phys. {\bf 6} (2004) 109
({\it http://stacks.iop.org/1367-2630/6/109});
Physica Scripta {\bf T121} (2005) 94 (hep-ph/0504110); 
S.~Pascoli and S.T.~Petcov, hep-ph/0308034 and 
arXiv:0711.   [hep-ph].
%
C.~Aalseth {\it et al.}, hep-ph/0412300;
A.~Morales and J.~Morales, 
{Nucl.\ Phys.\ B (Proc. Suppl.)} {\bf 114} (2003) 141.
%

\bibitem{SDavid07} S. Davidson {\it et al.}, 
Phys.\ Rev.\ Lett. {\bf 99} (2007) 161801.


\bibitem{rad1} J.~A.~Casas, J.~R.~Espinosa, A.~Ibarra and I.~Navarro,
  Nucl.\ Phys.\  B {\bf 573} (2000) 652;
T.~Miura, T.~Shindou and E.~Takasugi,
  Phys.\ Rev.\  D {\bf 66} (2002) 093002.
%

\bibitem{rad2} 
S.~Antusch {\it et al.}, Phys.\ Lett.\  B {\bf 519} (2001) 238.
%
\bibitem{PST06} 
S.~T.~Petcov, T.~Shindou and Y.~Takanishi,
  Nucl.\ Phys.\  B {\bf 738} (2006) 219.

\bibitem{SchValle80Doi81} J. Schechter and J.W.F. Valle, 
{Phys. Rev.} D {\bf 22} (1980) 2227;
M.~Doi {\it et al.},
{Phys. Lett.} B {\bf 102} (1981) 323.


\bibitem{BCGPRKL2} A. Bandyopadhyay {\it et al.}, 
{ Phys.\ Lett.} B {\bf 608} (2005) 115, and arXiv:0804.4857.


\bibitem{TSchwSNOW06} 
 T.~Schwetz, Phys.\ Scripta {\bf T127} (2006) 1.

\bibitem{Fogli06} G.L.~Fogli {\it et al.},
Prog. Part. Nucl. Phys. {\bf 57} (2006) 71.


\bibitem{TSchw08} T. Schwetz, M. Tortola and J.W.F. Valle, 
arXiv:0808.2016 [hep-ph].

\bibitem{CHOOZ} M. Apollonio {\it et al.}, 
                 Phys. Lett. B {\bf 466} (1999) 415.

\bibitem{Engelhard:2006yg}
G.~Engelhard, Y.~Grossman, E.~Nardi and Y.~Nir,
		Phys. Rev. Lett. {\bf 99} (2007) 081802.

\bibitem{IR041}  A.~Ibarra and G.G.~Ross,
Phys.\ Lett.\  B {\bf   591} (2004) 285;
P.H. Chankowski {\it et al.},
Nucl.\ Phys.\ B {\bf 690} (2004) 279. 

\bibitem{AIb06} A. Ibarra,  JHEP {\bf 0601} (2006) 064. 


\bibitem{FGY03} P.~H.~Frampton, S.~L.~Glashow and T.~Yanagida,
{Phys. Lett.} {\bf B548} (2002) 119;
A. Strumia and M. Raidal, Phys. Lett. B {\bf 553} (2003) 72.
%

\bibitem{DCHOOZ}  
F.~Ardellier {\it et al.}  [Double Chooz Collaboration],
  hep-ex/0606025.

\bibitem{DayaB} See, e.g., K. M. Heeger, talk given at 
Neutrino'06 International Conference,
June 13 - 19, 2006, Sant Fe, U.S.A.

\bibitem{Molinaro:2008cw}
  E.~Molinaro and S.~T.~Petcov,
  arXiv:0808.3534 [hep-ph].

\bibitem{SDavid08} S. Davidson, J. Garayoa, F. Palorini and N. Rius,
arXiv:0806.2832v3.


\end{thebibliography}
\end{document}